\documentclass[pra,twocolumn,showpacs,amsmath,amssymb,floatfix]
{revtex4}

\usepackage{graphicx}
\usepackage{bm}
\usepackage{psfrag}

\begin{document}

\title{Ground-state van der Waals forces in planar multilayer
magnetodielectrics}

\author{Stefan Yoshi Buhmann}

\author{Dirk-Gunnar Welsch}

\affiliation{Theoretisch-Physikalisches Institut,
Friedrich-Schiller-Universit\"{a}t Jena,
Max-Wien-Platz 1, 07743 Jena, Germany}

\author{Thomas Kampf}

\affiliation{Fachbereich Physik, Universit\"{a}t Rostock,
Universit\"{a}tsplatz 3, 18051 Rostock, Germany}

\date{\today}

\begin{abstract}
Within the frame of lowest-order perturbation theory, the van der
Waals potential of a ground-state atom placed within an arbitrary
dispersing and absorbing magnetodielectric multilayer system is given.
Examples of an atom situated in front of a magnetodielectric plate or
between two such plates are studied in detail. Special emphasis is
placed on the competing attractive and repulsive force components
associated with the electric and magnetic matter properties,
respectively, and conditions for the formation of repulsive potential
walls are given. Both numerical and analytical results are presented.
\end{abstract}


\maketitle

%%%%%%%%%%%%%%%%%%%%%%%%%%%%%%%%%%%%%%%%%%%%%%%%%%%%%%%%%%%%%%%%%%%%%%

\section{Introduction}
\label{sec1}

It is well known that an atom near a neutral macroscopic body is
subject to a force, even if the atom and the body are in the
(unpolarized) ground states. The existence of the force commonly
called van der Waals (vdW) force has been experimentally well
established. In particular, vdW forces on atoms in multilayer
systems have been observed via mechanical means in atomic beam
transmission \cite{Anderson88} and quantum reflection experiments 
\cite{Shimizu01}, and via spectroscopic means \cite{Sandoghdar92},
inter alia frequency modulated selective reflection spectroscopy
\cite{Oria91}. 

As long as the atom-body separation is sufficiently large compared
with the atomic radius on the one hand, and the typical distance
between the atomic constituents of the body on the other hand, the vdW
force can be calculated within the frame of macroscopic
electrodynamics. A unified theory that covers both the nonretarded
distance regime---already studied by Lennard-Jones in 1932
\cite{LennardJones32}---and the retarded distance regime was first 
given in 1948 by Casimir and Polder \cite{Casimir48}. With this in
mind, the force has also been called Casimir-Polder force. Casimir's
and Polder's theory is based on exact quantum electrodynamics (QED),
the electromagnetic field being quantized in terms of normal modes.
The coupling energy of a ground-state atom with the body-assisted
electromagnetic vacuum field is calculated in lowest order of
perturbation theory, and the vdW force emerges as the gradient of this
coupling energy---the vdW potential. This formalism first applied to
the case of an atom placed in front of a perfectly conducting plate
was later extended to excited atomic energy eigenstates
\cite{Barton74} as well as to an atom between two perfectly conducting
plates \cite{Walther97}. Moreover, the concept has been used to
calculate the vdW force acting on an atom placed in front of a
semi-infinite dielectric half space \cite{Tikochinski93} or between
two dielectric plates of finite thickness \cite{Zhou95}. Recently, the
ideas of Casimir and Polder have been generalized to allow for
dispersing and absorbing bodies \cite{Buhmann04a,Buhmann04b}, which
inhibit electromagnetic-field quantization via a standard normal-mode
expansion in general.

In parallel with the sophistication of Casimir's and Polder's concept
based on exact quantum electrodynamics, a semiphenomenological
approach to the problem of the vdW force has been established and
widely used. According to this approach, the coupling energy is
expressed in terms of correlation functions for the atom and/or the
electromagnetic field which in turn are related to susceptibilities
via the dissipation-fluctuation theorem. The result---which in
principle applies to arbitrary geometries---was first applied to a
ground-state atom placed in front of a perfectly conducting half space
\cite{McLachlan63}, a dielectric half space \cite{McLachlan63b}, and a
dielectric two-layer system \cite{Wylie84}. Later, atoms in excited
energy eigenstates were included in the concept \cite{Wylie85}.
Effects of surface roughness \cite{Henkel98}, finite temperature
\cite{McLachlan63c} and--- in the case of the semi-infinite half
space---different materials such as birefringent dielectric
\cite{Gorza01} and even magnetodielectric matter \cite{Kryszewki92}
have been considered. 

Apart from the two main routes outlined above, a number of other 
methods have been suggested and applied to various systems. The vdW
potential of a two-level atom in front of a perfectly conducting half
space has been derived upon using nonperturbative spectrum-summation
techniques \cite{Renne71} and classical electrodynamics with random
fluctuations \cite{Boyer72}. The problem of the vdW force acting on a
ground-state atom in front of a dielectric half space has been 
treated via microscopic models \cite{Renne71b}, $S$-matrix formalism
\cite{Mavroyannis63}, and source theory \cite{Schwinger78}. 
Electrostatic methods applicable to the nonretarded distance regime
have been used to determine the vdW force acting on an excited-state
atom in front of a semi-infinite half space filled by a birefringent
dielectric \cite{Fichet95}, and the problem of the vdW force acting
on an atom in front of a nondispersive dielectric three-layer system 
has been studied \cite{Courtois96}. Within the frame of macroscopic
quantum electrodynamics, a dynamical approach to the vdW force has
recently been developed in order to study time-dependent forces in the
case of atoms initially prepared in an arbitrary excited quantum state
\cite{Buhmann04b}.

In the large body of work on vdW forces and related electromagnetic
forces (such as the vdW force between two atoms or the Casimir force
between two macroscopic bodies) the electric properties of the
involved material objects have typically been the focus of interest.
Nevertheless, the interaction of objects possessing also noticeable
magnetic properties---a problem which has regained topicality due to
the recent fabrication of metamaterials with controllable
electromagnetic properties in the microwave regime
\cite{Pendry99,Smith00}---has been of interest. The fact that
Maxwell's equations in the absence of (free) charges and currents are
invariant under a duality transformation between electric and magnetic
fields can be exploited to extend the notion of forces acting on
electrically polarizable objects to objects with magnetic properties.
Thus, knowing the attractive vdW force between two electrically
polarizable particles (e.g., atoms), one can infer the existence of an
analogeous attractive force between two magnetically polarizable
particles, which may be obtained from the former by replacing the
electric polarizabilities by the corresponding magnetic ones. In
contrast, the force between two polarizable particles of opposed type
is repulsive \cite{Sucher70}. While the repulsive vdW potential in the
retarded limit obeys the same $1/r^7$ power law ($r$, distance between
the particles) as the attractive vdW potential (in the case of two
particles of the same type), but is smaller in magnitude than the
latter by a factor of $7/23$ \cite{Sucher68,Boyer69}, the leading
contribution to the repulsive vdW potential in the nonretarded limit
is proportional to $1/r^4$, which contrasts with the
$1/r^6$-dependence of the attractive vdW potential. This difference
can be understood by regarding the first particle as an oscillating
electric dipole creating an electromagnetic field which acts on the
second, electrically or magnetically polarizable particle. Due to the
fact that in the nonretarded limit the electromagnetic field is
dominated by the electrostatic field, the force on a electrically
polarizable particle is stronger than the force on a magnetically
polarizable one \cite{Farina02}.

Similar considerations can also be made for other systems. So, the
attractive Casimir force between two infinitely permeable plates 
corresponds to the force between two perfectly conducting plates
by virtue of duality, whereas the force between two plates of
different type is repulsive and smaller than the equal-type force by a
factor of $7/8$ \cite{Boyer74}. For realistic plates the situation
becomes more involved. In particular, the repulsive Casimir force
between a purely dielectric and a purely magnetic plate observed in
the retarded limit shows the same $1/z^4$ power law ($z$, separation
of the plates) as the attractive force between two dielectric plates,
whereas in the nonretarded limit the repulsive force behaves like
$1/z$ and the attractive force like $1/z^3$ \cite{Henkel04}. If one of
the plates is dielectric while the other one is magnetodielectric,
then the electric and magnetic properties of the second plate 
compete in determining the sign of the Casimir force
\cite{Kenneth02,Tomas04}.

It is known that the force acting on a magnetically polarizable 
particle in front of a perfectly conducting plate is repulsive in the
retarded limit \cite{Boyer69}. By virtue of duality a corresponding
repulsive force is expected to act on an electrically polarizable
particle such as a ground-state atom which is located in front of an
infinitely permeable plate. Thus the question arises what kind of
force could be observed in the case of a genuinely magnetodielectric
plate. Maybe the effect of a repulsive force component in such a
system is easier accessible to experimental verification than that of
a repulsive component of the Casimir force between two macroscopic
bodies, where force measurements are currently restricted to distance
regimes of purely attractive forces \cite{Ianuzzi03}. Moreover, the
recently reported production of metamaterials with controllable
magnetodielectric properties in the microwave regime
\cite{Pendry99,Smith00} opens the perspective of engineering vdW
potentials with desired properties.

In this paper we consider the vdW interaction of a ground-state atom
with planar, dispersing, and absorbing magnetodielectric bodies.
Starting from the general expression for the vdW potential in case of
an arbitrary planar multilayer system, as can be derived in lowest
order of perturbation theory within the frame of QED in linear, causal
media, we give a detailed analysis of the vdW potential of the atom
being located (i) in front of a magnetodielectric plate and (ii)
beween two magnetodielectric plates. In particular, we address the
question if and how the competition of electric and magnetic
properties of the material can give rise to a repulsive force. In this
context we study the influence of effects such as material absorption,
finite layer thickness, and multiple reflections. 

The paper is organized as follows. In Sec.~\ref{sec2} the vdW
potential of a ground-state atom in an arbitrary planar
magnetodielectric multilayer system is given. A detailed analysis of
typical examples is given in Sec.~\ref{sec3}, followed by a summary
and concluding remarks in Sec.~\ref{sec4}.

%%%%%%%%%%%%%%%%%%%%%%%%%%%%%%%%%%%%%%%%%%%%%%%%%%%%%%%%%%%%%%%%%%%%%%

\section{Basic equations}
\label{sec2}

Consider a neutral, nonpolar, ground-state atomic system such as an
atom or a molecule (briefly referred to as atom in the following) 
at position $\mathbf{r}_\mathrm{A}$ within an arbitrary arrangement of
linear magnetodielectric bodies, which is characterized by a
permittivity $\varepsilon(\mathbf{r},\omega)$ and a permeability
$\mu(\mathbf{r},\omega)$, which are spatially varying, complex-valued
functions of frequency, with the corresponding Kramers-Kronig
relations being satisfied. The position-dependent fluctuations of the
body-assisted electromagnetic field give rise to a force on the atom
which, within leading-or\-der perturbation theory, can be derived from
the vdW potential \cite{Buhmann04b}
\begin{equation}
\label{eq1}
     U(\mathbf{r}_\mathrm{A})
     = \frac{\hbar\mu_0}{2\pi}
     \int_0^{\infty} \mathrm{d} u \,u^2 \alpha^{(0)}(iu)
     \,\mathrm{Tr}\,
     \bm{G}^{(1)}(\mathbf{r}_\mathrm{A},\mathbf{r}_\mathrm{A},iu)
\end{equation}
according to
\begin{equation}
\label{eq2}
\mathbf{F}(\mathbf{r}_\mathrm{A})=
-\bm{\nabla}_{\!\!\mathrm{A}}U(\mathbf{r}_\mathrm{A})
\end{equation}
($\bm{\nabla}_{\!\!\mathrm{A}}$ $\!\equiv$
$\!\bm{\nabla}_{\!\mathbf{r}_\mathrm{A}}$). In Eq.~(\ref{eq1}), 
\begin{equation}
\label{eq3}
     \alpha^{(0)}(\omega)=\lim_{\epsilon\to 0}
     \frac{2}{3\hbar}\sum_k
     \frac{\omega_{k0}}{\omega_{k0}^2-\omega^2
     - i\omega\epsilon}
     \,|\mathbf{d}_{0k}|^2
\end{equation}
is the ground-state polarizability of the atom in lowest nonvanishing
order of perturbation theory [$\omega_{k0}$ $\!\equiv$ $\!(E_k$ $\!-$
$\!E_0)/\hbar$, (un\-per\-turbed) atomic transition frequencies;
\mbox{$\mathbf{d}_{0k}$ 
$\!\equiv$ $\!\langle 0|\hat{\mathbf{d}}|k\rangle$}, atomic
electric-dipole transition matrix elements], and
$\bm{G}^{(1)}(\mathbf{r},\mathbf{r}',iu)$ is the scattering part of
the classical Green tensor of the electromagnetic field,
\begin{equation}
\label{eq4}
\bm{G}(\mathbf{r},\mathbf{r}',\omega)
=\bm{G}^{(0)}(\mathbf{r},\mathbf{r}',\omega)
+\bm{G}^{(1)}(\mathbf{r},\mathbf{r}',\omega)
\end{equation}
[$\bm{G}^{(0)}(\mathbf{r},\mathbf{r}',\omega)$, bulk part], which is
the solution to the equation
\begin{equation}
\label{eq5}
      \left[
      \bm{\nabla}\times\kappa(\mathbf{r},\omega)\bm{\nabla}\times
      -\frac{\omega^2}{c^2}\,\varepsilon(\mathbf{r},\omega)
      \right]
      \bm{G}(\mathbf{r},\mathbf{r}',\omega)
      = \bm{\delta}(\mathbf{r}-\mathbf{r}')
\end{equation}
[$\kappa(\mathbf{r},\omega)$ $\!=$ $\!\mu^{-1}(\mathbf{r},\omega)$] 
together with the boundary condition 
\begin{equation}
\label{eq5.1}
\bm{G}(\mathbf{r},\mathbf{r}',\omega)\to 0  
\quad\mbox{for}\quad|\mathbf{r}-\mathbf{r}'|\to\infty.
\end{equation}

%%%%%%%%%%%%%%%  F I G U R E %%%%%%%%%%%%%%%%%%%%%%
\begin{figure}[!t!]
\noindent
\begin{center}
\includegraphics[width=\linewidth]{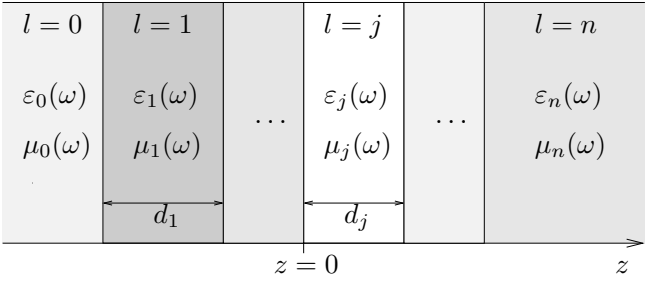}
\end{center}
\caption{
\label{fig1}
Sketch of the planar multilayer system.}
\end{figure}
%%%%%%%%%%%%%%%%%%%%%%%%%%%%%%%%%%%%%%%%%%%%%%%%%%%%%%%%%

In what follows we assume that the bodies surrounding the atom form a
planar multilayer system, i.e., a stack of $n$ $\!+$ $\!1$ layers 
labelled by $l$ ($l$ $\!=$ $0,\ldots,n$) of thicknesses $d_l$ 
with planar parallel boundary surfaces, where
$\varepsilon(\mathbf{r},\omega)$ $\!=$ $\!\varepsilon_l(\omega)$ and
$\mu(\mathbf{r},\omega)$ $\!=$ $\mu_l(\omega)$ in layer $l$. 
The coordinate system is chosen such that the layers are perpendicular
to the $z$ axis and extend from \mbox{$z$ $\!=$ $\!0$} to $z$ $\!=$
$\!d_l$ for $l$ $\!\neq$ $\!0,n$ and from \mbox{$z$ $\!=$ $\!0$} to
$z$ $\!=$ $\!-\infty$ ($\infty$) for $l$ $\!=$ $\!0$ ($n$), cf.
Fig.~\ref{fig1}. The scattering part of the Green tensor at imaginary
frequencies for $\mathbf{r}$ and $\mathbf{r}'$ in layer $j$ can be
given by \cite{Chew95}
\begin{equation}
\label{eq6}
\bm{G}^{(1)}(\mathbf{r},\mathbf{r}',iu) =
\int\mathrm{d}^2q\,
e^{i\mathbf{q}\cdot(\mathbf{r}-\mathbf{r}')}
\bm{G}^{(1)}(\mathbf{q},z,z',iu)
\end{equation}
($\mathbf{q}\perp\mathbf{e}_z$). Here,
\begin{align}
\label{eq7}
&\bm{G}^{(1)}(\mathbf{q},z,z',iu) = \frac{\mu_j(iu)}{8\pi^2b_j}
 \sum_{\sigma=s,p}\biggl\{ \frac{r^\sigma_{j-}r^\sigma_{j+}
 e^{-2b_jd_j}}{D_j^\sigma}\nonumber\\
&\quad\times\,\Bigl[\mathbf{e}_\sigma^+\mathbf{e}_\sigma^+ 
 e^{-b_j(z-z')}
 +\mathbf{e}_\sigma^-\mathbf{e}_\sigma^-
 e^{b_j(z-z')}\Bigr]\nonumber\\
&\qquad+\frac{1}{D_j^\sigma}
 \Bigl[\mathbf{e}_\sigma^+\mathbf{e}_\sigma^- r^\sigma_{j-} 
 e^{-b_j(z+z')}\nonumber\\
&\qquad\qquad\quad+\mathbf{e}_\sigma^-\mathbf{e}_\sigma^+
r^\sigma_{j+}
 e^{-2b_jd_j}e^{b_j(z+z')}\Bigr]\biggr\}
\end{align}
for $j$ $\!>$ $\!0$, where
\begin{equation}
\label{eq9}
\mathbf{e}_s^\pm=\mathbf{e}_q\times\mathbf{e}_z,
 \quad\mathbf{e}_p^\pm=-\frac{1}{k_j}(iq\mathbf{e}_z
 \pm b_j\mathbf{e}_q)
\end{equation}
($\mathbf{e}_q$ $\!=$ $\!\mathbf{q}/q$, $q$ $\!=$ $\!|\mathbf{q}|$)
with
\begin{equation}
\label{eq10}
k_j=\frac{u}{c}\sqrt{\varepsilon_j(iu)\mu_j(iu)}
\end{equation}
are the polarization vectors for $s$- and $p$-polarized waves
propagating in the positive/negative $z$-direction, $r^\sigma_{j-}$
and $r^\sigma_{j+}$ are the generalized coefficients for reflection at
the left/right boundary of layer $j$, which can be calculated with
the aid of the recursive relations
\begin{alignat}{1}
\label{eq11}
&r^s_{l\pm}=
\frac{\left(\frac{\mu_{l\pm 1}}{b_{l\pm 1}}-\frac{\mu_l}{b_l}\right)
+\left(\frac{\mu_{l\pm 1}}{b_{l\pm 1}}+\frac{\mu_l}{b_l}\right)
e^{-2b_{l\pm 1}d_{l\pm 1}}r^s_{l\pm 1\pm}}
{\left(\frac{\mu_{l\pm 1}}{b_{l\pm 1}}+\frac{\mu_l}{b_l}\right)
+\left(\frac{\mu_{l\pm 1}}{b_{l\pm 1}}-\frac{\mu_l}{b_l}\right)
e^{-2b_{l\pm 1}d_{l\pm 1}}r^s_{l\pm 1\pm}}\, ,\\
\label{eq12}
&r^p_{l\pm}
=\frac{\left(\frac{\varepsilon_{l\pm 1}}{b_{l\pm 1}}
-\frac{\varepsilon_l}{b_l}\right)
+\left(\frac{\varepsilon_{l\pm 1}}{b_{l\pm 1}}
+\frac{\varepsilon_l}{b_l}\right)
e^{-2b_{l\pm 1}d_{l\pm 1}}r^p_{l\pm 1\pm}}
{\left(\frac{\varepsilon_{l\pm 1}}{b_{l\pm 1}}
+\frac{\varepsilon_l}{b_l}\right)
+\left(\frac{\varepsilon_{l\pm 1}}{b_{l\pm 1}}
-\frac{\varepsilon_l}{b_l}\right)
e^{-2b_{l\pm 1}d_{l\pm 1}}r^p_{l\pm 1\pm}}
\end{alignat}
($l$ $\!=$ $\!1,\ldots,j$ for $r^\sigma_{l-}$, 
$l$ $\!=$ $\!j,\ldots,n$ $\!-$ $\!1$ for $r^\sigma_{l+}$,
\mbox{$r^\sigma_{0-}$ $\!=$ $\!r^\sigma_{n+}$ $\!=$ $\!0$}),
\begin{equation}
\label{eq8}
b_l = \sqrt{\frac{u^2}{c^2}\ \varepsilon_l(iu)\mu_l(iu) + q^2}
\end{equation}
is the imaginary part of the $z$-component of the wave vector in layer
$l$, and finally
\begin{equation}
\label{eq15}
D_j^\sigma=1-r_{j-}^\sigma r_{j+}^\sigma e^{-2b_jd_j}.
\end{equation}

Let the atom be situated in the otherwise empty layer $j$, i.e., 
$\varepsilon_j(iu)$ $\!=$ $\!\mu_j(iu)$ $\!\equiv$ $\!1$ and
\begin{equation}
\label{eq16}
b_j=\sqrt{\frac{u^2}{c^2}+q^2} \equiv b.
\end{equation}
To calculate the vdW potential, we substitute Eq.~(\ref{eq6}) together
with Eq.~(\ref{eq7}) into Eq.~(\ref{eq1}), thereby omitting the
irrelevant position-independent terms. Evaluating the trace with the
aid of the relations
\begin{align}
\label{eq17}
&\mathbf{e}_s^\pm\cdot\mathbf{e}_s^\pm
=\mathbf{e}_s^\pm\cdot\mathbf{e}_s^\mp=1,\\
\label{eq17b}
&\mathbf{e}_p^\pm\cdot\mathbf{e}_p^\pm=1, 
\quad \mathbf{e}_p^\pm\cdot\mathbf{e}_p^\mp
=-1-2\left(\frac{qc}{u}\right)^2,
\end{align}
which directly follow from Eqs.~(\ref{eq9}), (\ref{eq10}) and (\ref{eq8}), we
realize that the resulting integrand of the $\mathbf{q}$-integral only
depends on $q$. Thus after introducing polar coordinates in the
$q_xq_y$-plane, we can easily perform the angular integration, leading
to
\begin{align}
\label{eq18}
&U(z_\mathrm{A}) = \frac{\hbar\mu_0}{8\pi^2}
 \int_0^{\infty} \mathrm{d} u \,u^2 \alpha^{(0)}(iu)
 \int_0^\infty\mathrm{d}q\,\frac{q}{b}\nonumber\\
&\quad\times\,\Biggl\{
 e^{-2bz_\mathrm{A}}\biggl[\frac{r_{j-}^s}{D_j^s}
 -\biggl(1+2\frac{q^2c^2}{u^2}\biggr)
 \frac{r_{j-}^p}{D_j^p}\biggr]\nonumber\\
&\qquad\quad+e^{-2b(d_j-z_\mathrm{A})}
 \biggl[\frac{r_{j+}^s}{D_j^s}-\biggl(1+2\frac{q^2c^2}{u^2}\biggr)
 \frac{r_{j+}^p}{D_j^p}\biggr]\Biggr\}.
\end{align}
Note that Eq.~(\ref{eq7}) and thus Eq.~(\ref{eq18}) also apply to
the case $j$ $\!=$ $\!0$ if $d_0$ is formally set equal to zero
\mbox{($d_0$ $\!\equiv$ $\!0$)}.

Equation (\ref{eq18}) together with Eq.~(\ref{eq3}) and
Eqs.~(\ref{eq11})--(\ref{eq16}) presents the vdW potential of a
ground-state atom within a general planar magnetodielectric multilayer
system. Note that instead of calculating the generalized reflection
coefficients $r_{j\pm}^\sigma$ from the permittivities and
permeabilities of the individual layers via
Eqs.~(\ref{eq11})--(\ref{eq8}) (as we shall do in this paper), it is
possible to determine them experimentally by appropriate reflectivity
measurements (cf., e.g., Ref.~\cite{Thakur04}). In the case where the
atom is placed (in free space) in front of the multilayer system ($j$
$\!=$ $\!n$), Eq.~(\ref{eq18}) reduces to
\begin{align}
\label{eq19}
U(z_\mathrm{A}) 
& = \frac{\hbar\mu_0}{8\pi^2}
 \int_0^{\infty} \mathrm{d} u \,u^2 \alpha^{(0)}(iu)
 \int_0^\infty\mathrm{d}q\,
 \frac{q}{b}e^{-2bz_\mathrm{A}}\\\nonumber
&\qquad\qquad\times\,\biggl[r_{n-}^s
-\biggl(1+2\frac{q^2c^2}{u^2}\biggr)r_{n-}^p\biggr].
\end{align}

%%%%%%%%%%%%%%%%%%%%%%%%%%%%%%%%%%%%%%%%%%%%%%%%%%%%%%%%%%%%%%%%%%%%%%

\section{Specific examples}
\label{sec3}

Typical features of the vdW potential of an atom in the case of
magnetodielectric multilayer systems---in particular the competing
influence of the electric and magnetic properties of the layers, the
effect of material absorption, the influence of finite layer thickness
or multiple reflections---can be already illustrated by studying 
relatively simple systems consisting of only a few layers. 

%%%%%%%%%%%%%%%%%%%%%%%%%%%%%%%%%%%%%%%%%%%%%%%%%%%%%%%%%%%%%%%%%%%%%%

\subsection{Perfectly reflecting plate}
\label{sec3.0}

As a preliminary investigation, let us consider the idealizing 
case of an atom positioned in the $n$th (empty) layer in front of a 
perfectly reflecting (multilayer) plate, i.e., $|r_{n-}^s|$ $\!=$
$\!|r_{n-}^p|$ $\!=$ $\!1$. We begin with the case 
\begin{equation}
\label{eq19-1}
r_{n-}^s = -1, \quad r_{n-}^p = +1, 
\end{equation}
which corresponds to the limit of a perfectly conducting
plate $\varepsilon_{n-1}$ $\!\to$ $\!\infty$, as can be seen from 
Eqs.~(\ref{eq11}) and (\ref{eq12}) [together with
Eq.~(\ref{eq8})]. Changing the integration variables in
Eq.~(\ref{eq19}) according to $(u,q)$ $\!\mapsto$ $(u,b)$, we obtain
the attractive potential
\begin{align}
\label{eq20}
U(z_\mathrm{A})
& = -\frac{\hbar}{4\pi^2\varepsilon_0}
 \int_0^{\infty} \mathrm{d} u\,\alpha^{(0)}(iu)
 \int_{u/c}^\infty\mathrm{d}b\,
 b^2e^{-2bz_\mathrm{A}}\nonumber\\
& = -\frac{\hbar}{16\pi^2\varepsilon_0z_\mathrm{A}^3}
 \int_0^{\infty} \mathrm{d} u \, \alpha^{(0)}(iu)
 e^{-2uz_\mathrm{A}/c}\nonumber\\
&\qquad\quad\times\left[1+2\left(\frac{uz_\mathrm{A}}{c}\right)
 +2\left(\frac{uz_\mathrm{A}}{c}\right)^2\right],
\end{align}
which is exactly the result found by Casimir and Polder 
for the potential of a ground-state atom in front of 
a perfectly conducting plate \cite{Casimir48}. In the long-distance 
(i.e., retarded) limit, \mbox{$z_\mathrm{A}$ $\!\gg$
$c/\omega_\mathrm{A}^-$} [$\omega_\mathrm{A}^-$ $\!=$
$\!\mathrm{min}(\{\omega_{k0}|k$ $\!=$ $\!1,2,\ldots\})$], the atomic
polarizability $\alpha^{(0)}(iu)$ may be approximately replaced with
its static value $\alpha^{(0)}(0)$ and put in front of the integral,
leading to
\begin{equation}
\label{eq21}
U(z_\mathrm{A})=
-\frac{3\hbar c\alpha^{(0)}(0)}{32\pi^2\varepsilon_0z_\mathrm{A}^4}\,.
\end{equation}
In the short-distance (i.e., nonretarded) limit, \mbox{$z_\mathrm{A}$
$\!\ll$ $\!c/\omega_\mathrm{A}^+$} [$\omega_\mathrm{A}^+$ $\!=$
$\!\mathrm{max}(\{\omega_{k0}|k$ $\!=$ $\!1,2,\ldots\})$], we may
approximately set \mbox{$e^{-2uz_\mathrm{A}/c}$ $\!=1$} in
Eq.~(\ref{eq20}) and neglect the second and third terms in the square
brackets to recover, on recalling Eq.~(\ref{eq3}), the result of
Lennard-Jones \cite{LennardJones32},
\begin{equation}
\label{eq22}
U(z_\mathrm{A})=
-\frac{1}{48\pi\varepsilon_0z_\mathrm{A}^3}\sum_k
   |\mathbf{d}_{0k}|^2
= -\frac{\langle 0|\hat{\mathbf{d}}^2|0\rangle}
{48\pi\varepsilon_0z_\mathrm{A}^3}\,.
\end{equation}

In contrast, if the layer facing the atom is supposed to be infinitely
permeable, i.e., $\mu_{n-1}\! \to\! \infty$, Eqs.~(\ref{eq11}) and
(\ref{eq12}) [together with Eq.~(\ref{eq8})] lead to 
\begin{equation}
\label{eq20-1}
r_{n-}^s = 1, \quad r_{n-}^p = -1,
\end{equation}
and Eq.~(\ref{eq19}) yields the repulsive potential
\begin{align}
\label{eq22.1}
U(z_\mathrm{A})
& = \frac{\hbar}{16\pi^2\varepsilon_0z_\mathrm{A}^3}
 \int_0^{\infty} \mathrm{d} u \, \alpha^{(0)}(iu)
 e^{-2uz_\mathrm{A}/c}\nonumber\\
&\qquad\times\left[1+2\left(\frac{uz_\mathrm{A}}{c}\right)
 +2\left(\frac{uz_\mathrm{A}}{c}\right)^2\right].
\end{align}
In particular, in the long-distance limit we have 
[cf. Eq.~(\ref{eq21})]
\begin{equation}
\label{eq22.1-1}
U(z_\mathrm{A})=
\frac{3\hbar c\alpha^{(0)}(0)}{32\pi^2\varepsilon_0z_\mathrm{A}^4}\,,
\end{equation}
which by means of a duality transformation [$\alpha^{(0)}(0)$ 
$\!\equiv$ $\alpha^{(0)}_\mathrm{e}(0)$ $\!\mapsto$ 
$\!\alpha^{(0)}_\mathrm{m}(0)$] can be transformed to the result 
obtained in Ref.~\cite{Boyer69} for a magnetically polarizable
particle [of polarizability $\alpha^{(0)}_\mathrm{m}(0)$]  
in front of a perfectly conducting plate.  
Application of the duality transformation to Eq.~(\ref{eq22.1})
generalizes the result in Ref.~\cite{Boyer69} to arbitrary distances. 

%%%%%%%%%%%%%%%%%%%%%%%%%%%%%%%%%%%%%%%%%%%%%%%%%%%%%%%%%%%%%%%%%%%%%%

\subsection{Infinitely thick plate}
\label{sec3.1}

To be more realistic, let us first consider an atom in front of a 
sufficiently thick magnetodielectric plate which may be modelled by
a semi-infinite half space [$n$ $\!=$ $\!j$ $\!=$ $\!1$, 
$\varepsilon_1(\omega)$ $\!=$ $\!\mu_1(\omega)$ $\!\equiv$ $\!1$,
$\varepsilon_0(\omega)$ $\!\equiv$ $\!\varepsilon(\omega)$,
$\mu_0(\omega)$ $\!\equiv$ $\!\mu(\omega)$]. Substituting the
reflection coefficients as follow from Eqs.~(\ref{eq11}) and
(\ref{eq12}) into Eq.~(\ref{eq19}), we find, on recalling
Eq.~(\ref{eq8}), that ($b_0$ $\!\equiv$ $\!b_\mathrm{M}$)
\begin{align}
\label{eq23}
&U(z_{\rm A})
 = \frac{\hbar\mu_0}{8\pi^2}
 \int_0^{\infty} \mathrm{d} u \,u^2 \alpha^{(0)}(iu)
 \int_0^\infty\mathrm{d}q\,\frac{q}{b}e^{-2bz_\mathrm{A}}\nonumber\\
&\quad\times \biggl[\frac{\mu(iu)b-b_\mathrm{M}}
 {\mu(iu)b+b_\mathrm{M}}
 -\biggl(1+2\frac{q^2c^2}{u^2}\biggr)
 \frac{\varepsilon(iu)b-b_\mathrm{M}}
 {\varepsilon(iu)b+b_\mathrm{M}}\biggr].
\end{align}
Equation (\ref{eq23}) is equivalent to the result derived in
Ref.~\cite{Kryszewki92} semiphenomenologically within the frame of
linear response theory. Note that the concept of linear response
theory may render erroneous results when trying to go beyond
perturbation theory (cf. the remark in Ref.~\cite{Buhmann04b}).

To further evaluate Eq.~(\ref{eq23}), let us model the permittivity
and (paramagnetric) permeability, respectively, by
\begin{equation}
\label{eq31}
\varepsilon(\omega) = 1+\,\frac{\omega_\mathrm{Pe}^2}
{\omega_\mathrm{Te}^2-\omega^2-i\omega\gamma_\mathrm{e}}
\end{equation}
and
\begin{equation}
\label{eq32}
\mu(\omega) = 1 +\,\frac{\omega_\mathrm{Pm}^2}
{\omega_\mathrm{Tm}^2-\omega^2-i\omega\gamma_\mathrm{m}}\,.
\end{equation} 
It can then be shown that in the long-distance limit, i.e.,
\mbox{$z_\mathrm{A}$ $\!\gg\!$ $c/\omega_\mathrm{A}^-$},
$z_\mathrm{A}$ $\!\gg\!$ $c/\omega_\mathrm{M}^-$ 
[$\omega_\mathrm{M}^-$ $\!=\!$
$\mathrm{min}(\omega_\mathrm{Te},\omega_\mathrm{Tm})$], 
Eq.~(\ref{eq23}) reduces to (see Appendix \ref{appA})
\begin{equation}
\label{eq24}
U(z_\mathrm{A})=\frac{C_4}{z_\mathrm{A}^4}\,,
\end{equation}
where
\begin{align}
\label{eq24.1}
C_4
=& -\frac{3\hbar c\alpha^{(0)}(0)}{64\pi^2\varepsilon_0}
\int_{1}^\infty\mathrm{d} v\, 
\left[\left(\frac{2}{v^2}-\frac{1}{v^4}\right)\right.\nonumber\\
&\quad\times\,\frac{\varepsilon(0)v-\sqrt{\varepsilon(0)\mu(0)-1 + v^2 }}
{\varepsilon(0)v+\sqrt{\varepsilon(0)\mu(0)- 1 + v^2 }}
\nonumber\\
&\qquad-\frac{1}{v^4}\,\left.
\frac{\mu(0)v-\sqrt{\varepsilon(0)\mu(0)-1 + v^2}}
{\mu(0)v+\sqrt{\varepsilon(0)\mu(0)-1 + v^2 }}\right],
\end{align}
while in the short-distance limit, i.e.,
\mbox{$z_\mathrm{A}$ $\!\ll$ $\!c/[\omega_\mathrm{A}^+n(0)]$} and/or
\mbox{$z_\mathrm{A}$ $\!\ll$ $\!c/[\omega_\mathrm{M}^+n(0)]$} 
[$\omega_\mathrm{M}^+$ $\!=\!$ 
$\mathrm{max}(\omega_\mathrm{Te},\omega_\mathrm{Tm})$,
$n(0)$ $\!=$ $\!\sqrt{\varepsilon(0)\mu(0)}$],
Eq.~(\ref{eq23}) leads to (see Appendix \ref{appA})
\begin{equation}
\label{eq25}
U(z_\mathrm{A})= -\ \frac{C_3}{z_\mathrm{A}^3}
 +\frac{C_1}{z_\mathrm{A}}\,,
\end{equation}
where
\begin{equation}
\label{eq25.1}
C_3
 = \frac{\hbar}{16\pi^2\varepsilon_0}
 \int_0^\infty\mathrm{d}u\ \alpha^{(0)}(iu)
\frac{\varepsilon(iu)-1}{\varepsilon(iu)+1}\ge 0
\end{equation}
and
\begin{align}
\label{eq25.2}
C_1 &= \frac{\mu_0\hbar}{16\pi^2}
 \int_0^\infty\mathrm{d}u\ u^2\alpha^{(0)}(iu)
 \biggl\{\frac{\varepsilon(iu)-1}{\varepsilon(iu)+1}\nonumber\\
&\quad+ \frac{\mu(iu)-1}{\mu(iu)+1}
 + \frac{2\varepsilon(iu)[\varepsilon(iu)\mu(iu)-1]}
 {[\varepsilon(iu)+1]^2}
 \biggr\}\ge 0. 
\end{align}
We have numerically checked the asymptotic behaviour given by
Eqs.~(\ref{eq24})--(\ref{eq25.2}) for the case of a two-level atom.
{F}rom the derivation it is clear that Eqs.~(\ref{eq24}) and
(\ref{eq25}) also remain valid when---in generalization of
Eqs.~(\ref{eq31}) and (\ref{eq32}), respectively---more than one
matter resonance is taken into account. Needless to say that the
minimum $\omega_\mathrm{M}^-$ and the maximum $\omega_\mathrm{M}^+$
are then defined with respect to all matter resonances. 

\begin{figure}[!t!]
\noindent
\begin{center}
\includegraphics[width=\linewidth]{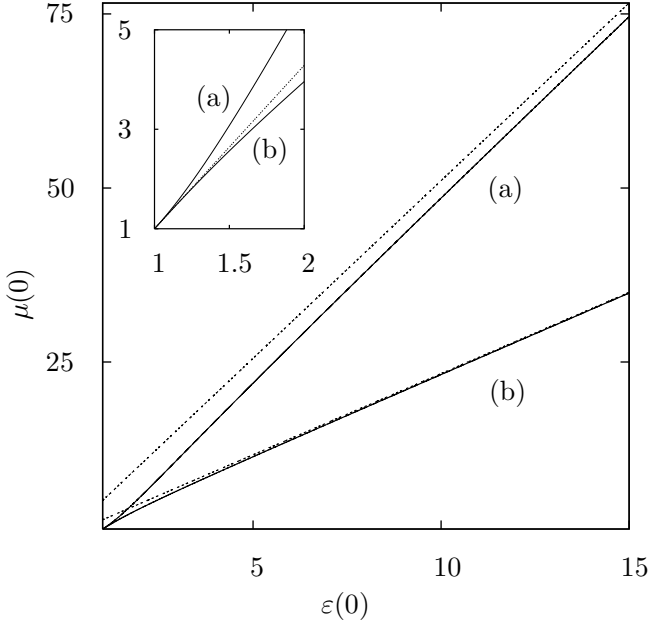}
\end{center}
\caption{
\label{fig5}
Border between attractive and repulsive long-distance vdW potentials
of an atom in front of (a) a thick  and (b) a thin magnetodielectric
plate according to Eqs.~(\ref{eq24.1}) ($C_4$ $\!=$ $\!0$) and
(\ref{eq32.1}) ($D_5$ $\!=$ $\!0$). The broken curves show the
asymptotic behaviour as given by Eqs.~(\ref{eq26.4}) (inset) and
(\ref{eq26.5.2}) in case (a) and by Eqs.~(\ref{eq33}) (inset) and
(\ref{eq34}) in case (b). 
}
\end{figure}

Inspection of Eq.~(\ref{eq24.1}) reveals that the coefficient $C_4$ 
in Eq.~(\ref{eq24}) for the long-distance behaviour of the vdW
potential is negative (positive) for a purely electric (magnetic)
plate, corresponding to an attractive (repulsive) force. For a
genuinely magnetodielectric plate the situation is more involved.
As the coefficient $C_4$ monotoneously decreases with increasing
$\varepsilon(0)$ and monotoneously increases with increasing $\mu(0)$,
\begin{equation}
\label{eq26.1}
\frac{\partial C_4}{\partial\varepsilon(0)}<0, \quad
\frac{\partial C_4}{\partial\mu(0)}>0, \quad
\end{equation}
the border between the attractive and repulsive potential, i.e., $C_4$
$\!=$ $\!0$, can be marked by a unique curve in the
$\varepsilon(0)\mu(0)$-plane [curves (a) in Fig.~\ref{fig5}]. 
In particular, in the limits of weak and strong magnetodielectric
properties the integral in Eq.~(\ref{eq24.1}) can be evaluated
analytically. For weak magnetodielectric properties, i.e.,
\mbox{$\chi_\mathrm{e}(0)$ $\!\equiv$ $\!\varepsilon(0)$ $\!-1$
$\!\ll$ $\!1$} and $\chi_\mathrm{m}(0)$ $\!\equiv$ $\!\mu(0)$ $\!-$
$\!1$
$\!\ll$ $\!1$, the linear expansions
\begin{align}
\label{eq26.2}
&\frac{\varepsilon(0)v-\sqrt{\varepsilon(0)\mu(0)-1 + v^2 }}
{\varepsilon(0)v+\sqrt{\varepsilon(0)\mu(0)- 1 + v^2 }}
\nonumber\\
&\qquad
\simeq
\left[\frac{1}{2}-\frac{1}{4v^2}\right]\chi_\mathrm{e}(0)
-\frac{1}{4v^2}\chi_\mathrm{m}(0)
\end{align}
and
\begin{align}
\label{eq26.3}
&\frac{\mu(0)v-\sqrt{\varepsilon(0)\mu(0)-1 + v^2}}
{\mu(0)v+\sqrt{\varepsilon(0)\mu(0)-1 + v^2 }}
\nonumber\\
&\qquad
\simeq
-\frac{1}{4v^2}\chi_\mathrm{e}(0)
+\left[\frac{1}{2}-\frac{1}{4v^2}\right]\chi_\mathrm{m}(0)
\end{align}
lead to
\begin{equation}
\label{eq26.4}
C_4=-\frac{\hbar c\alpha^{(0)}(0)}{640\pi^2\varepsilon_0}
\bigl[23\ \chi_\mathrm{e}(0)-7\chi_\mathrm{m}(0)\bigr].
\end{equation}
For strong magnetodielectric properties, i.e., $\varepsilon(0)$
$\!\gg$ $\!1$ and $\mu(0)$ $\!\gg$ $\!1$, we may approximately set, on
noting that large values of $v$ are effectively suppressed in the
integral in Eq.~(\ref{eq24.1}),
\begin{equation}
\label{eq26.5}
\sqrt{\varepsilon(0)\mu(0)-1 + v^2 }
\simeq
\sqrt{\varepsilon(0)\mu(0)}\,,
\end{equation}
thus
\begin{align}
\label{eq26.5.2}
C_4 =& -\frac{3\hbar c\alpha^{(0)}(0)}{64\pi^2\varepsilon_0}
 \biggl[-\,\frac{2}{Z^3}\mathrm{ln}(1\!+\!Z)
 +\frac{2}{Z^2}+\frac{4}{Z}\mathrm{ln}(1\!+\!Z)\nonumber\\
& - \frac{1}{Z}-\frac{4}{3}-Z+2Z^2-2Z^3
 \mathrm{ln}\biggl(1+\frac{1}{Z}\biggr)\biggr],
\end{align}
where $Z$ $\!\equiv$ $\!\sqrt{\mu(0)/\varepsilon(0)}$ is the static
impedance of the material. Setting $C_4$ $\!=$ $\!0$ in
Eqs.~(\ref{eq26.4}) and (\ref{eq26.5.2}), we obtain the asymptotic
behaviour of the border curve in the two limiting cases. In
particular, from Eq.~(\ref{eq26.5.2}) it follows that $Z$ $\!=$
$\!2.26$. In conclusion one can say that in the long-distance limit a
repulsive vdW potential can be realized if the static magnetic
properties are stronger than the static electric properties,
\mbox{$\chi_\mathrm{m}(0)/\chi_\mathrm{e}(0)$ $\!\ge$ $\!23/7$ 
$\!=$ $\!3.29$} for 
weak magnetodielectric properties, and \mbox{$\mu(0)/\varepsilon(0)$ 
$\!\ge$ $\!5.11$} for strong magnetodielectric properties. 
%%%%%%%%%%%%%%%  F I G U R E %%%%%%%%%%%%%%%%%%%%%%%%%%%%%%%%%%%%%%%%%
%%%%%%%%%%%%%%%%%%%%%%%%%%%%%%%%%%%%%%%%%%%%%%%%%%%%%%%%%%%%%%%%%%%%%%

Apart from the different distance laws, the short-distance vdW
potential, Eq.~(\ref{eq25}), differs from the long-distance potential,
Eq.~(\ref{eq24}), in two respects. First, the relevant coefficients
$C_3$ and $C_1$ are not only determined by the static values of the
permittivity and the permeability, as is seen from Eqs.~(\ref{eq25.1})
and (\ref{eq25.2}), and second, Eqs.~(\ref{eq25})--(\ref{eq25.2})
reveal that electric and magnetic properties give rise to 
potentials with different distance laws and signs [$C_3$ $\!>$ $\!0$ 
dominant (and $C_1$ $\!>$ $\!0$) if \mbox{$\varepsilon$ $\!\neq$
$\!1$} and $\mu$ $\!=$ $\!1$, while \mbox{$C_3$ $\!=$ $\!0$} and $C_1$
$\!>$ $\!0$ if \mbox{$\varepsilon$ $\!=$ $\!1$} and $\mu$ $\!\neq$
$\!1$]. However, although for the case of a purely magnetic plate a
repulsive vdW potential proportional to $1/z_\mathrm{A}$ is predicted,
in practice the attractive $1/z_\mathrm{A}^3$ term will always
dominate for sufficiently small values of $z_\mathrm{A}$, because of
the always existing electric properties of the plate. Hence when in
the long-distance limit the potential becomes repulsive due to
sufficiently strong magnetic properties, then the formation of a
potential wall at intermediate distances becomes possible. It is
evident that with decreasing strength of the electric properties the
maximum of the wall is shifted to smaller distances while increasing
in height. 

In the limiting case of weak electric properties, i.e.,
$\omega_\mathrm{Pe}/\omega_\mathrm{Te}$ $\!\ll$ $\!1$
and  $\omega_\mathrm{Pe}/\omega_\mathrm{Pm}$ $\!\ll$ $\!1$
[recall Eqs.~(\ref{eq31}) and (\ref{eq32})] one can thus expect that
the wall is situated within the short-distance range, so that
Eqs.~(\ref{eq25})--(\ref{eq25.2}) can be used to determine both its
position and height. {F}rom Eq.~(\ref{eq25}) we find that the wall
maximum is at 
\begin{equation}
\label{eq26.6}
z_\mathrm{A}^\mathrm{max}=\sqrt{\frac{3C_3}{C_1}}
\end{equation}
and has a height of
\begin{equation}
\label{eq26.7}
U(z_\mathrm{A}^\mathrm{max}) 
 =\frac{2}{3}\sqrt{\frac{C_1^3}{3C_3}}\, .
\end{equation}
In order to estimate the integrals in Eqs.~(\ref{eq25.1}) and
(\ref{eq25.2}) for the coefficients $C_3$ and $C_1$, respectively, let
us restrict our attention to the case of a two-level atom and
disregard absorption ($\gamma_\mathrm{e}$ $\!\simeq$ $\!0$, 
$\!\gamma_\mathrm{m}$ $\!\simeq$ $\!0$). Straightforward calculation
then yields ($\omega_\mathrm{Pe}/\omega_\mathrm{Te}$ $\!\ll$
$\!1$, $\omega_\mathrm{Pe}/\omega_\mathrm{Pm}$ $\!\ll$ $\!1$)
\begin{equation}
\label{eq26.8}
C_3\simeq\frac{|\mathbf{d}_{01}|^2}{96\pi\varepsilon_0}\,
 \frac{\omega_\mathrm{Pe}^2}{\omega_\mathrm{Te}^2}\,
 \frac{\omega_\mathrm{Te}}{\omega_{10}\!+\!\omega_\mathrm{Te}}
\end{equation} 
and
\begin{align}
\label{eq26.9}
C_1 \simeq&  \,\frac{\mu_0\hbar}{16\pi^2}
 \!\int_0^\infty\!\mathrm{d}u\, u^2\alpha^{(0)}(iu)
 \!\left[ \frac{\mu(iu)\!-\!1}{\mu(iu)\!+\!1} 
+ \frac{\mu(iu)\!-\!1}{2}\right]\nonumber\\
= & \,\frac{\mu_0|\mathbf{d}_{01}|^2\omega_\mathrm{Pm}^2}{96\pi}\,
 \frac{\omega_\mathrm{10}
 (2\omega_\mathrm{10}+\omega_\mathrm{Sm}+\omega_\mathrm{Tm})}
 {(\omega_\mathrm{10}+\omega_\mathrm{Sm})
 (\omega_\mathrm{10}+\omega_\mathrm{Tm})}
\end{align}
[$\omega_\mathrm{Sm}$ $\!=\!$ $\!(\omega_\mathrm{Tm}^2$ $\!+$
$\!\frac{1}{2}\omega_\mathrm{Pm}^2)^{1/2}$]. Substitution of
Eqs.~(\ref{eq26.8}) and (\ref{eq26.9}) into Eqs.~(\ref{eq26.6}) and
(\ref{eq26.7}), respectively, eventually leads to 
\begin{align}
\label{eq26.10}
z_\mathrm{A}^\mathrm{max} =&\, 
 \frac{c}{\omega_\mathrm{Pm}}
 \frac{\omega_\mathrm{Pe}}{\omega_\mathrm{Te}}
 \sqrt{\frac{\omega_\mathrm{Te}(\omega_{10}+\omega_\mathrm{Tm})}
 {\omega_{10}(\omega_{10}+\omega_\mathrm{Te})}}
 \nonumber\\
&\quad\times\,\sqrt{\frac
 {3(\omega_{10}\!+\!\omega_\mathrm{Sm})}
 {(2\omega_\mathrm{10}+\omega_\mathrm{Sm}
 +\omega_\mathrm{Tm})}}
\end{align}
and
\begin{align}
\label{eq26.12}
U(z_\mathrm{A}^\mathrm{max}) 
=& \,\frac{|\mathbf{d}_{01}|^2
 \omega_\mathrm{Pm}^3}
 {48\pi \varepsilon_0 c^3}\,
 \frac{\omega_\mathrm{Te}}{\omega_\mathrm{Pe}}
 \sqrt{\frac{\omega_{10}+\omega_\mathrm{Te}}
 {\omega_\mathrm{Te}}}
 \nonumber\\
&\quad\times\,\left[\frac{\omega_{10}(2\omega_{10}
 \!+\!\omega_\mathrm{Sm}+\omega_\mathrm{Tm})}
 {3(\omega_{10}+\omega_\mathrm{Sm})
 (\omega_{10}+\omega_\mathrm{Tm})}\right]^{\frac{3}{2}}\, .
\end{align}
Note that consistency with the assumption of the wall being observed
at short distances requires that $z_\mathrm{A}^\mathrm{max}$ $\!\ll$
$c/\omega_\mathrm{M}^+$---a condition which can be easily fulfilled for
sufficiently small values of $\omega_\mathrm{Pe}/\omega_\mathrm{Pm}$.
Inspection of Eq.~(\ref{eq26.12}) shows that the height of the wall
increases with increasing $\omega_\mathrm{Pm}$, decreasing
$\omega_\mathrm{Tm}$, and decreasing
$\omega_\mathrm{Pe}/\omega_\mathrm{Te}$ $\!=$
$\!\sqrt{\varepsilon(0)-1}$. Since the dependence of
$U(z_\mathrm{A}^\mathrm{max})$ on $\omega_\mathrm{Pm}$ is 
much stronger than its dependendence on $\omega_\mathrm{Tm}$,
the wall height increases with $\omega_\mathrm{Tm}$ for given
$\omega_\mathrm{Pm}/\omega_\mathrm{Tm}$ $\!=$ $\!\sqrt{\mu(0)-1}$.

%%%%%%%%%%%%%%%  F I G U R E %%%%%%%%%%%%%%%%%%%%%%%%%%%%%%%%%%%%%%%%%
\begin{figure}[!t!]
\noindent
\begin{center}
\includegraphics[width=\linewidth]{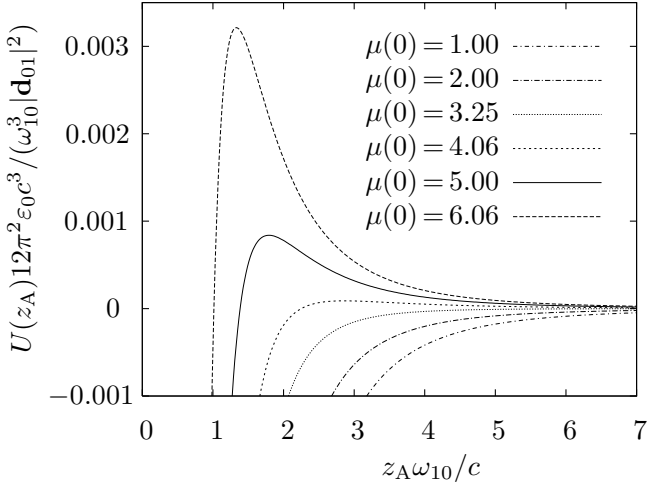}
\end{center}
\caption{
\label{fig2}
The vdW potential of a ground-state two-level atom situated in front
of an infinitely thick magnetodielectric plate is shown as a function
of the distance between the atom and the plate for different values of
$\mu(0)$ ($\omega_\mathrm{Pe}/\omega_{10}$ $\!=$ $\!0.75$,
$\omega_\mathrm{Te}/\omega_{10}$ $\!=$ $\!1.03$,
$\omega_\mathrm{Tm}/\omega_{10}$ $\!=$ $\!1$,
$\gamma_\mathrm{e}/\omega_{10}$ 
$\!=$ $\!\gamma_\mathrm{m}/\omega_{10}$ $\!=$ $\!0.001$).
}
\end{figure}
%%%%%%%%%%%%%%%%%%%%%%%%%%%%%%%%%%%%%%%%%%%%%%%%%%%%%%%%%%%%%%%%%%%%%%
%%%%%%%%%%%%%%%  F I G U R E %%%%%%%%%%%%%%%%%%%%%%%%%%%%%%%%%%%%%%%%%
\begin{figure}[!t!]
\noindent
\begin{center}
\includegraphics[width=\linewidth]{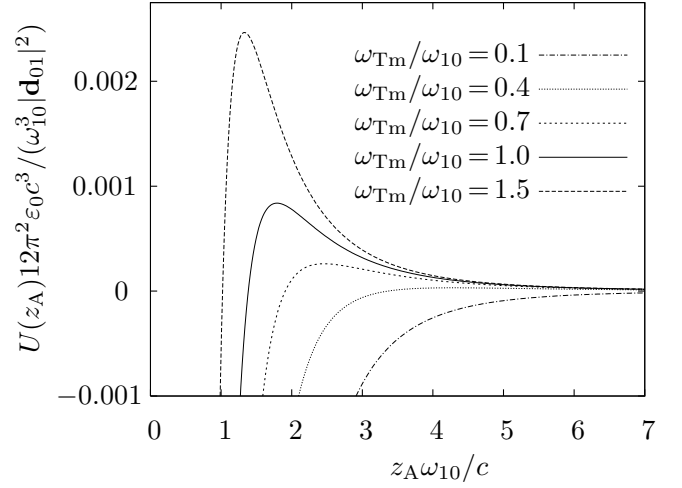}
\end{center}
\caption{
\label{fig3}
The vdW potential of a ground-state two-level atom situated in front
of an infinitely thick magnetodielectric plate is shown as a function
of the distance between the atom and the plate for different values of
$\omega_\mathrm{Tm}$ [$\mu(0)$ $\!=$ $\!5$,
$\omega_\mathrm{Pe}/\omega_{10}$ $\!=$ $\!0.75$,
$\omega_\mathrm{Te}/\omega_{10}$ $\!=$ $\!1.03$,
$\gamma_\mathrm{e}/\omega_{10}$ 
$\!=$ $\!\gamma_\mathrm{m}/\omega_{10}$ $\!=$ $\!0.001$].
}
\end{figure}
%%%%%%%%%%%%%%%%%%%%%%%%%%%%%%%%%%%%%%%%%%%%%%%%%%%%%%%%%%%%%%%%%%%%%%

The distance dependence of the vdW potential, as calculated
from Eq.~(\ref{eq23}) for a two-level atom in front of a thick
magnetodielectric plate whose permittivity and permeability are
modelled by Eqs.~(\ref{eq31}) and (\ref{eq32}), respectively, is
illustrated in Figs.~\ref{fig2} and \ref{fig3}. The figures reveal
that the results derived above for the case where the potential wall
is observed in the short-distance range remain qualitatively valid
also for larger distances. So, from Fig.~\ref{fig2} it is seen that,
for chosen values of $\omega_\mathrm{Tm}$ and $\gamma_\mathrm{m}$,
the potential wall begins to form and grows in height as $\mu(0)$
increases, while Fig.~\ref{fig3} confirms that, for chosen values of
$\mu(0)$ and $\gamma_\mathrm{m}$, the height of the wall increases
with $\omega_\mathrm{Tm}$. In conclusion one can say that the
formation of a noticeable potential wall requires materials whose
static permeability substantially exceeds the static permittivity,
thereby featuring magnetic resonance frequencies as high as possible.

\begin{figure}[!t!]
\noindent
\begin{center}
\includegraphics[width=\linewidth]{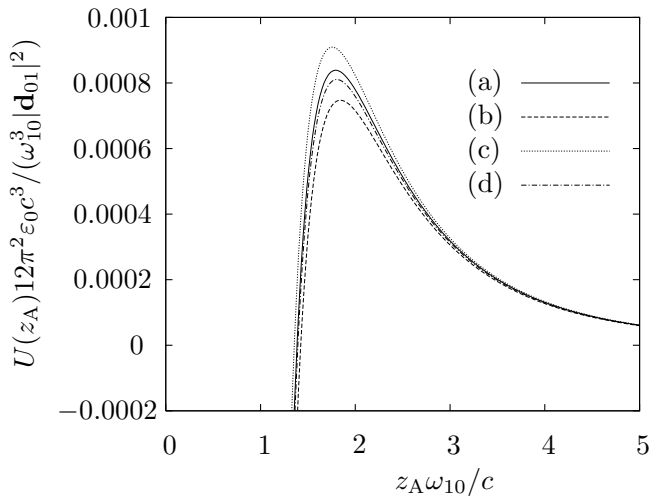}
\end{center}
\caption{
\label{fig4}
The vdW potential of a ground-state two-level atom situated in front
of a infinitely thick magnetodielectric plate is shown as a function
of the distance between the atom and the plate for different values of
the  absorption parameters:
(a) $\gamma_\mathrm{e}/\omega_{10}$ $\!=$ $\!0.001$, 
$\gamma_\mathrm{m}/\omega_{10}$ $\!=$ $\!0.001$,
(b) $\gamma_\mathrm{e}/\omega_{10}$ $\!=$ $\!0.001$, 
$\gamma_\mathrm{m}/\omega_{10}$ $\!=$ $\!0.05$,
(c) $\gamma_\mathrm{e}/\omega_{10}$ $\!=$ $\!0.05$, 
$\gamma_\mathrm{m}/\omega_{10}$ $\!=$ $\!0.001$,
(d) $\gamma_\mathrm{e}/\omega_{10}$ $\!=$ $\!0.05$, 
$\gamma_\mathrm{m}/\omega_{10}$ $\!=$ $\!0.05$
($\omega_\mathrm{Pe}/\omega_{10}$ $\!=$ $\!0.75$,
$\omega_\mathrm{Te}/\omega_{10}$ $\!=$ $\!1.03$,
$\omega_\mathrm{Pm}/\omega_{10}$ $\!=$ $\!2$,
$\omega_\mathrm{Tm}/\omega_{10}$ $\!=$ $\!1$).
}
\end{figure}

To study the dependence of the vdW potential on material absorption as
characterized by the parameters $\gamma_\mathrm{e}$ and
$\gamma_\mathrm{m}$ in Eqs.~(\ref{eq31}) and (\ref{eq32}), we first
consider the limiting behaviour of the potential for long and short
distances. As the potential in the long-distance limit can be given in
terms of the static permittivity and permeability, which do not depend
on the absorption parameters, material absorption has no influence on
the vdW force for asymptotically large distances. In contrast, 
absorption can affect the potential in the short-distance
limit. {F}rom Eqs.~(\ref{eq31}) and (\ref{eq32}) the inequalities 
\begin{equation}
\label{eq26.13}
\frac{\partial\varepsilon(iu)}{\partial\gamma_\mathrm{e}}<0, \quad
\frac{\partial\mu(iu)}{\partial\gamma_\mathrm{m}}<0
\end{equation}
are seen to be valid. Combining them with Eqs.~(\ref{eq25.1}) and
(\ref{eq25.2}) reveals that
\begin{eqnarray}
\label{eq26.14}
\frac{\partial C_3}{\partial\gamma_\mathrm{e}}<0, \quad
\frac{\partial C_3}{\partial\gamma_\mathrm{m}}=0,\\
\label{eq26.14-1}
\frac{\partial C_1}{\partial\gamma_\mathrm{e}}<0, \quad
\frac{\partial C_1}{\partial\gamma_\mathrm{m}}<0.
\end{eqnarray}
Provided that the magnetic properties of the medium are sufficiently
strong to support the formation of a potential wall, these
inequalities imply [cf. Eq.~(\ref{eq25})] that increasing 
$\gamma_\mathrm{e}$ ($\gamma_\mathrm{m}$) leads to a shift of the wall
towards smaller (larger) distances, while increasing (decreasing) its
height. Thus an increase of $\gamma_\mathrm{e}$ yields a stronger
repulsive potential, whereas a simultaneous increase of both
absorption parameters is expected to lead to a reduction of the wall
height in general. This behaviour is confirmed by the examples shown
in Fig.~\ref{fig4}, where the vdW potential of a two-level atom as
given by Eq.~(\ref{eq23}) is displayed as a function of the distance
between the atom and the plate for different values of the two
absorption parameters. Note the reduced influence of absorption at
large distances---in agreement with the arguments given above. 
%%%%%%%%%%%%%%%  F I G U R E %%%%%%%%%%%%%%%%%%%%%%%%%%%%%%%%%%%%%%%%%
%%%%%%%%%%%%%%%%%%%%%%%%%%%%%%%%%%%%%%%%%%%%%%%%%%%%%%%%%%%%%%%%%%%%%%

In view of left-handed materials (see, e.g.,
Refs.~\cite{Pendry99,Smith00,Veselago68}), which simultaneously
exhibit negative real parts of $\varepsilon(\omega)$ and $\mu(\omega)$
within some (real) frequency interval such that the real part of the
refractive index becomes negative therein, the question may arise
whether these materials would have an exceptional effect on the
ground-state vdW force. The answer is obviously no, because the
ground-state vdW potential as given by Eq.~(\ref{eq23}) is expressed
in terms of the always positive values of the permittivity and the
permeability at imaginary frequencies. Clearly, the situation may
change for an atom prepared in an excited state. In such a case, the
vdW potential is essentially determined by the real part of the Green
tensor taken at frequencies close to the transition frequencies of the
atom \cite{Buhmann04b}. When there are transition frequencies that lie
in frequency intervals where the material behaves left-handed, then
particularities may occur.

%%%%%%%%%%%%%%%%%%%%%%%%%%%%%%%%%%%%%%%%%%%%%%%%%%%%%%%%%%%%%%%%%%%%%%

\subsection{Plate of finite thickness}
\label{sec3.2}

Let us now consider an atom in front of a magnetodielectric plate
of finite thickness $d_1$ $\!\equiv\!$ $d$
[$n$ $\!=$ $\!j$ $\!=$ $\!2$, $\varepsilon_1(\omega)$ $\!\equiv$ 
$\!\varepsilon(\omega)$, $\mu_1(\omega)$ $\!\equiv$ $\!\mu(\omega)$,
$\varepsilon_0(\omega)$ $\!=$ $\!\varepsilon_2(\omega)$ $\!\equiv$ 
$\!1$, $\mu_0(\omega)$ $\!=$ $\!\mu_2(\omega)$ $\!\equiv$ $\!1$].
Substituting the reflection coefficients calculated from
Eqs.~(\ref{eq11}) and (\ref{eq12}) into Eq.~(\ref{eq19}), we
derive ($b_1$ $\!\equiv$ $\!b_\mathrm{M}$)
\begin{align}
\label{eq27}
&U(z_\mathrm{A})
=\frac{\hbar\mu_0}{8\pi^2}
 \int_0^{\infty} \mathrm{d} u \,u^2 \alpha^{(0)}(iu)
 \int_0^\infty\mathrm{d}q\,\frac{q}{b}\,e^{-2bz_\mathrm{A}}
 \nonumber\\&\hspace{1ex}
\times\,
 \biggl\{-\biggl(1+2\frac{q^2c^2}{u^2}\biggr)
 \nonumber\\&\hspace{4ex}
\times\,
 \frac{[\varepsilon^2(iu)b^2-b_\mathrm{M}^2]\tanh(b_\mathrm{M}d)}
 {2\varepsilon(iu)b b_\mathrm{M}+
 [\varepsilon^2(iu)b^2+b_\mathrm{M}^2]\tanh(b_\mathrm{M}d)}
 \nonumber\\&\hspace{6ex}
+\frac{[\mu^2(iu)b^2-b_\mathrm{M}^2]\tanh(b_\mathrm{M}d)}
 {2\mu(iu)b b_\mathrm{M}+[\mu^2(iu)b^2+b_\mathrm{M}^2]
 \tanh(b_\mathrm{M}d)}\biggr\}.
\end{align}
It is obvious that the integration in Eq.~(\ref{eq27}) is effectively
limited by the exponential factor $e^{-2bz_\mathrm{A}}$ to a circular
region where \mbox{$b$ $\!\lesssim$ $\!1/(2z_\mathrm{A})$}. In 
particular, in the limit of a sufficiently thick plate,
\mbox{$d$ $\!\gg$ $\!z_\mathrm{A}$}, the estimate
\begin{equation}
\label{eq27a}
b_\mathrm{M}d\ge bd\sim\frac{d}{2z_\mathrm{A}}\gg 1
\end{equation}
[recall Eqs.~(\ref{eq8}) and (\ref{eq16})] is approximately
valid within (the major part of) the effective region of integration,
and one may hence make the approximation $\tanh(b_\mathrm{M}d)$
$\!\simeq$ $\!1$ in Eq.~(\ref{eq27}), which obviously leads back
to Eq.~(\ref{eq23}) valid for an infinitely thick plate. On the
contrary, in the limit of an asymptotically thin plate, $n(0)d$
$\!\ll$ $\!z_\mathrm{A}$, we find that the inequalities
\begin{align}
\label{eq27b}
b_\mathrm{M}d &\le \sqrt{\varepsilon(iu)\mu(iu)}\,bd
\le\sqrt{\varepsilon(0)\mu(0)}\,bd
\nonumber\\
&\le \frac{n(0)d}{2z_\mathrm{A}} \ll 1
\end{align}
hold in the effective region of integration, and one may hence perform
a linear expansion of the integrand in Eq.~(\ref{eq27}) in terms of
$b_\mathrm{M}d$, resulting in 
\begin{align}
\label{eq27c}
&U(z_\mathrm{A})
=\frac{\hbar\mu_0d}{8\pi^2}
 \int_0^{\infty} \mathrm{d} u \,u^2 \alpha^{(0)}(iu)
 \int_0^\infty\mathrm{d}q\,\frac{q}{b}\,
 e^{-2bz_\mathrm{A}}
 \nonumber\\&\hspace{1ex}
\times\,
 \biggl[-\biggl(1+2\frac{q^2c^2}{u^2}\biggr)
 \frac{\varepsilon^2(iu)b^2-b_\mathrm{M}^2}
 {2\varepsilon(iu)b}
+\frac{\mu^2(iu)b^2-b_\mathrm{M}^2}
 {2\mu(iu)b}\biggr].
\end{align}

\begin{figure}[!t!]
\noindent
\begin{center}
\includegraphics[width=\linewidth]{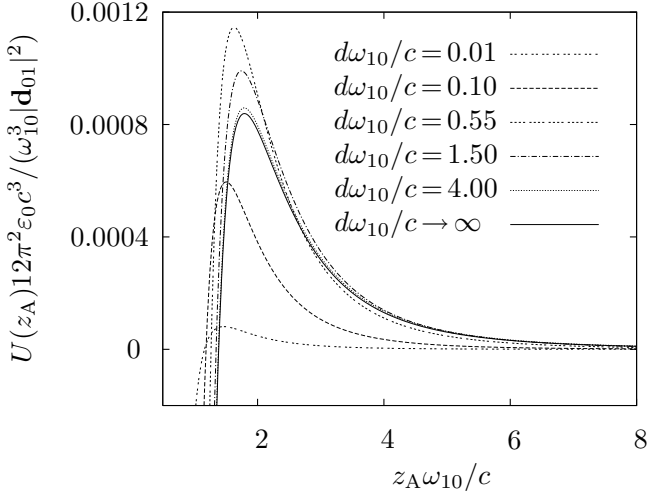}
\end{center}
\caption{
\label{fig6}
The vdW energy of a ground-state two-level atom situated in front of
a magnetodielectric plate is shown as a function of the distance
between the body and the interface for different values of the plate
thickness $d$ 
($\omega_\mathrm{Pe}/\omega_{10}$ $\!=$ $\!0.75$,
$\omega_\mathrm{Te}/\omega_{10}$ $\!=$ $\!1.03$,
$\omega_\mathrm{Pm}/\omega_{10}$ $\!=$ $\!2$,
$\omega_\mathrm{Tm}/\omega_{10}$ $\!=$ $\!1$,
$\gamma_\mathrm{e}/\omega_{10}$ 
$\!=$ $\!\gamma_\mathrm{m}/\omega_{10}$ $\!=$ $\!0.001$).
}
\end{figure}

Provided that the magnetic properties are sufficiently strong, the
formation of a repulsive potential wall can be also observed in the
case of a genuinely magnetodielectric plate of finite thickness. 
Typical examples of the vdW potential obtained by numerical evaluation
of Eq.~(\ref{eq27}) for a two-level atom are shown in Fig.~\ref{fig6}. 
In the figure, the medium parameters correspond to those which have
been found in Sec.~\ref{sec3.1} to support the formation of a
potential wall in the case of an infinitely thick plate. We see that
the qualitative behaviour of the vdW potential is independent of the
plate thickness. In particular, all curves in Fig.~\ref{fig6} 
feature a repulsive long-range potential that leads to a potential
wall of finite height, the potential becoming attractive at very short
distances. However, the position and height of the wall are seen to
vary with the thickness of the plate. While the position of the wall
shifts only slightly as the plate thickness is changed from very small
to very large values, the height of the wall reacts very sensitively
as the plate thickness is varied. For small values of the thickness
the potential height is very small, it increases towards a maximum,
and then decreases asymptotically towards the value found for the
infinitely thick plate as the thickness is increased further towards
very large values. It is worth noting that there is an optimal plate
thickness for creating a maximum potential wall. In this case the
magnitude of the plate thickness is comparable to the position of the
potential maximum---a case which is realized between the two extremes
of infinitely thick and infinitely thin plates.
%%%%%%%%%%%%%%%  F I G U R E %%%%%%%%%%%%%%%%%%%%%%%%%%%%%%%%%%%%%%%%%
%%%%%%%%%%%%%%%%%%%%%%%%%%%%%%%%%%%%%%%%%%%%%%%%%%%%%%%%%%%%%%%%%%%%%%

In order to gain further insight into the competing electric and
magnetic effects on the formation of an potential wall, let us study
the case of an asymptotically thin plate as described by
Eq.~(\ref{eq27c}) in more detail and compare it with the case
of an infinitely thick plate studied in Sec.~\ref{sec3.1}. 
In the long-distance limit, \mbox{$z_\mathrm{A}$ $\!\gg$
$\!c/\omega_\mathrm{A}^-,\,c/\omega_\mathrm{M}^-$},
Eq.~(\ref{eq27c}) reduces to 
(see Appendix \ref{appA}) 
\begin{equation}
\label{eq28}
U(z_\mathrm{A})=\frac{D_5}{z_\mathrm{A}^5}\, ,
\end{equation} 
where
\begin{equation}
\label{eq28.1}
D_5=-\frac{\hbar c\alpha^{(0)}(0)d}{160\pi^2\varepsilon_0}\,
 \biggl[\frac{14\varepsilon^2(0)-9}{\varepsilon(0)}
 -\frac{6\mu^2(0)-1}{\mu(0)}\biggr]\,,
\end{equation}
while in the short-distance limit, $z_\mathrm{A}$ $\!\ll$
$\!c/[\omega_\mathrm{A}^+n(0)]$ and/or \mbox{$z_\mathrm{A}$ $\!\ll$
$\!c/[\omega_\mathrm{M}^+n(0)]$}, Eq.~(\ref{eq27c}) can be
approximated by (see Appendix \ref{appA}) 
\begin{equation}
\label{eq29}
U(z_\mathrm{A})=-\frac{D_4}{z_\mathrm{A}^4}
 +\frac{D_2}{z_\mathrm{A}^2}\, ,
\end{equation}
where
\begin{equation}
\label{eq29.1}
D_4= \frac{3\hbar d}{64\pi^2\varepsilon_0}
 \int_0^\infty\mathrm{d}u\,
 \alpha^{(0)}(iu)\frac{\varepsilon^2(iu)-1}{\varepsilon(iu)}\ge 0
\end{equation}
and
\begin{align}
\label{eq30}
&D_2 = \frac{\mu_0\hbar d}{64\pi^2}
 \int_0^\infty\mathrm{d}u\,u^2
 \alpha^{(0)}(iu)\Biggl\{\frac{\varepsilon^2(iu)-1}{\varepsilon(iu)}
 \nonumber\\&\hspace{6ex}
+\frac{\mu^2(iu)-1}{\mu(iu)}
 +\frac{2[\varepsilon(iu)\mu(iu)-1]}{\varepsilon(iu)}\Biggr\}\ge 0\,.
\end{align}
In the case of an asymptotically thin plate the border between
attractive and repulsive potentials is determined by the equation
\mbox{$D_5$ $\!=$ $\!0$}, because Eq.~(\ref{eq28.1}) reveals that
\begin{equation}
\label{eq31.1}
\frac{\partial D_5}{\partial\varepsilon(0)}<0, \quad
\frac{\partial D_5}{\partial\mu(0)}>0 
\end{equation}
[cf. Eq.~(\ref{eq26.1}) valid for an infinitely thick plate]. Since
for an asymptotically thin plate---in contrast to the infinitely thick
plate---the influence of electric and magnetic properties can be
completely separated into a sum of two terms, the equation $D_5$ $\!=$
$\!0$ can be solved analytically, leading to 
\begin{equation}
\label{eq32.1}
\mu(0)=\frac{14\varepsilon^2(0)-9+
\sqrt{196\varepsilon^4(0)-228\varepsilon^2(0)+81}}
{12\varepsilon(0)}
\end{equation}
[curves (b) in Fig.~\ref{fig5}]. For suff\-iciently weak
magnetodielectric properties, i.e., \mbox{$\chi_e(0)$ $\!\equiv$
$\!\varepsilon(0)$ $\!-$ $\!1$ $\!\ll$ $\!1$}, $\chi_m(0)$ $\!\equiv$
$\!\mu(0)$ $\!-$ $\!1$ $\!\ll$ $\!1$, a linear expansion of the
right-hand side of Eq.~(\ref{eq32.1}) reveals that a repulsive vdW
potential can be realized if the static magnetic properties are
stronger than the static electric properties by a factor
$\chi_m(0)/\chi_e(0)$ $\!\ge$ $\!23/7$ $\!=$ $\!3.29$, 
corresponding to 
\begin{equation}
\label{eq33}
D_5=-\frac{\hbar c\alpha^{(0)}(0)d}{160\pi^2\varepsilon_0}\,
 \bigl[23\chi_e(0)-7\chi_m(0)\bigr],
\end{equation}
as can be seen by linearly expanding the right-hand side of
Eq.~(\ref{eq28.1}). By comparing Eqs.~(\ref{eq26.4}) and (\ref{eq33})
we realize that in the limit of weak magnetodielectric properties the
border between attractive and the repulsive vdW potentials is the same
for the infinitely thick plate and the asymptotically thin plate (cf.
the inset in Fig.~\ref{fig5}). This result is an immediate consequence
of the fact that in this case the thick-plate potential is a linear
superposition of thin-plate potentials [see Sec.~\ref{sec3.3},
Eq.~(\ref{eq03})]. For strong magnetodielectric properties,
$\varepsilon(0)\!\gg\!1$, $\mu(0)\!\gg\!1$, the asymptotic behaviour
of the right-hand side of Eq.~(\ref{eq32.1}) shows that the vdW
potential becomes repulsive if \mbox{$\mu(0)/\varepsilon(0)$ $\!\ge$
$\!7/3$ $\!=$ $\!2.33$}, corresponding to 
\begin{equation}
\label{eq34}
D_5=-\frac{\hbar c\alpha^{(0)}(0)d}{80\pi^2\varepsilon_0}\,
 \bigl[7\varepsilon(0)-3\mu(0)\bigr],
\end{equation}
which follows from the corresponding asymptotic expansion of the
right-hand side of Eq.~(\ref{eq28.1}). Hence the region in the
$\varepsilon(0)\mu(0)$-plane that corresponds to a repulsive vdW force
is slightly increased in comparison to the infinitely thick plate.

As in the case of an infinitely thick plate, the electric and
magnetic properties of the medium give rise to competing attractive
and repulsive potential components, where again the attractive
potential component resulting from the electric properties dominates
in the limit $z_\mathrm{A}$ $\!\to$ $\!0$ [see
Eqs.~(\ref{eq29})--(\ref{eq30})]. This implies that for weak electric
properties ($\omega_\mathrm{Pe}/\omega_\mathrm{Te}$ $\!\ll$ $\!1$,
$\omega_\mathrm{Pe}/\omega_\mathrm{Pm}$ $\!\ll$ $\!1$) a 
potential wall is formed in the short-distance range. {F}rom
Eq.~(\ref{eq29}) it then follows that the wall is situated at 
\begin{equation}
\label{eq35}
z_\mathrm{A}^\mathrm{max}=\sqrt{\frac{2D_4}{D_2}}
\end{equation}
and has a height of
\begin{equation}
\label{eq36}
U(z_\mathrm{A}^\mathrm{max}) 
 =\frac{D_2^2}{4D_4}\,.
\end{equation}
For a two-level atom interacting with an almost nonabsorbing 
($\gamma_\mathrm{e}$ $\!\simeq$ $\!0$, $\gamma_\mathrm{m}$ $\!\simeq$
$\!0$) single-resonance medium exhibiting weak electric properties
($\omega_\mathrm{Pe}/\omega_\mathrm{Te}$ $\!\ll$ $\!1$,
$\omega_\mathrm{Pe}/\omega_\mathrm{Pm}$ $\!\ll$ $\!1$), the
coefficients $D_4$, Eq.~(\ref{eq29.1}), and $D_2$, Eq.~(\ref{eq30}),
can be evaluated according to
\begin{equation}
\label{eq37}
D_4=\frac{d|\mathbf{d}_{01}|^2}{32\pi\varepsilon_0}\,
 \frac{\omega_\mathrm{Pe}^2}{\omega_\mathrm{Te}^2}\,
 \frac{\omega_\mathrm{Te}}{\omega_{10}\!+\!\omega_\mathrm{Te}}
\end{equation} 
and
\begin{align}
\label{eq38}
D_2 &\simeq\frac{\mu_0\hbar d}{64\pi^2}
 \int_0^\infty\mathrm{d}u\, u^2\alpha^{(0)}(iu)
\nonumber\\&\hspace{4ex}\times\,
 \biggl[ \frac{\mu^2(iu)-1}{\mu(iu)} +\ 2\mu(iu)-2\biggr]
\nonumber\\
&=\frac{d\mu_0|\mathbf{d}_{01}|^2\omega_\mathrm{Pm}^2}{96\pi}\,
 \frac{\omega_\mathrm{10}(4\omega_\mathrm{10}\!+\!3\omega_\mathrm{Lm}
 \!+\!\omega_\mathrm{Tm})}
 {2(\omega_\mathrm{10}\!+\!\omega_\mathrm{Lm})
 (\omega_\mathrm{10}\!+\!\omega_\mathrm{Tm})}
\end{align}
($\omega_\mathrm{Lm}$ $\!\equiv$
$\!\sqrt{\omega_\mathrm{Tm}^2+\omega_\mathrm{Pm}^2}$). Substitution of
Eqs.~(\ref{eq37}) and (\ref{eq38}) into Eqs.~(\ref{eq35}) and
(\ref{eq36}) leads to
\begin{align}
\label{eq39}
&z_\mathrm{A}^\mathrm{max} =
 \frac{c}{\omega_\mathrm{Pm}}
 \frac{\omega_\mathrm{Pe}}{\omega_\mathrm{Te}}
 \sqrt{\frac{\omega_\mathrm{Te}
 (\omega_{10}+\omega_\mathrm{Tm})}
 {\omega_{10}(\omega_{10}+\omega_\mathrm{Te})}}
 \nonumber\\
&\hspace{12ex}\times\,
 \sqrt{\frac{12(\omega_{10}+\omega_\mathrm{Lm})}
 {4\omega_\mathrm{10}+3\omega_\mathrm{Lm}
 +\omega_\mathrm{Tm}}}
\end{align}
(with the consistency requirement $z_\mathrm{A}^\mathrm{max}$
$\!\ll$ $\!c/\omega_\mathrm{M}^+$ being fulfilled for sufficiently
small values of $\omega_\mathrm{Pe}/\omega_\mathrm{Pm}$) and
\begin{align}
\label{eq40}
U(z_\mathrm{A}^\mathrm{max}) 
& = \frac{d|\mathbf{d}_{01}|^2
 \omega_\mathrm{Pm}^4}
 {1152\pi \varepsilon_0 c^4}\,
 \frac{\omega_\mathrm{Te}^2}{\omega_\mathrm{Pe}^2}
 \frac{\omega_{10}+\omega_\mathrm{Te}}
 {\omega_\mathrm{Te}}
\nonumber\\
&\hspace{6ex}\times\,\biggl[\frac{\omega_{10}
 (4\omega_\mathrm{10}+3\omega_\mathrm{Lm}
 +\omega_\mathrm{Tm})}
 {2(\omega_{10}+\omega_\mathrm{Lm})
 (\omega_{10}+\omega_\mathrm{Tm})}\biggr]^2.
\end{align}
Comparing Eqs.~(\ref{eq39}) and (\ref{eq40}) with Eqs.~(\ref{eq26.10})
and (\ref{eq26.12}) valid for an infinitely thick plate, we find that
the dependence of both the position and the height of the potential
wall on the material parameters is very similar, so that the criteria
for having a noticeable potential wall given below Eq.~(\ref{eq26.12})
also apply to the case of an asymptotically thin plate. {F}rom the
result that the position of the wall is almost the same in both cases
it may be expected that the wall position slowly varies with the plate
thickness, which is in full agreement with the exact results in
Fig.~\ref{fig6}. Further, the height of the wall is---in agreement
with Fig.~\ref{fig6}---considerably smaller for the asmyptotically
thin plate. This can be seen by applying
\mbox{$d/z_\mathrm{A}^\mathrm{max}$ $\!\le$
$\!\sqrt{\varepsilon(0)\mu(0)}d/z_\mathrm{A}^\mathrm{max}$ $\!\ll$
$\!1$} together with Eq.~(\ref{eq39}) in Eq.~(\ref{eq40}),
leading to
\begin{align}
\label{eq40.1}
U(z_\mathrm{A}^\mathrm{max}) 
& \ll\frac{3|\mathbf{d}_{01}|^2\omega_\mathrm{Pm}^3}
 {768\pi \varepsilon_0 c^3}\,
 \frac{\omega_\mathrm{Te}}{\omega_\mathrm{Pe}}
 \sqrt{\frac{\omega_{10}\!+\!\omega_\mathrm{Te}}
 {\omega_\mathrm{Te}}}
\nonumber\\
&\hspace{6ex}\times\,\biggl[\frac{\omega_{10}
(4\omega_\mathrm{10}+3\omega_\mathrm{Lm}
 +\omega_\mathrm{Tm})}
 {3(\omega_{10}+\omega_\mathrm{Lm})
 (\omega_{10}+\omega_\mathrm{Tm})}\biggr]^{\frac{3}{2}} ,
\end{align}
the right-hand side of which is comparable to the right-hand side of
Eq.~(\ref{eq26.12}). Recall that the wall height does not monotonously
increase with the plate thickness in general [see Fig.~\ref{fig6}], as
could be expected from comparing the two limiting cases.  

%%%%%%%%%%%%%%%%%%%%%%%%%%%%%%%%%%%%%%%%%%%%%%%%%%%%%%%%%%%%%%%%%%%%%%

\subsection{Power laws and medium-assisted correlations}
\label{sec3.3}

Comparing the asymptotic power laws (\ref{eq24}) and (\ref{eq25})
obtained for an infinitely thick plate with those obtained for an
asymptotically thin plate, Eqs.~(\ref{eq28}) and (\ref{eq29}), we see
that in the latter case the powers of $1/z_\mathrm{A}$ are universally
increased by one. In both cases the long-distance vdW potential
follows a power law that is independent of the material properties of
the plate, the sign being determined by the relative strengths of
magnetic and electric properties (a purely electric plate creates an
attractive vdW potential, while a purely magnetic plate gives rise to
a repulsive one). Further, the short-distance results for plates of
different material properties differ in both sign and leading power
law (the repulsive potential created by a purely magnetic plate being
weaker than the attractive potential created by a purely electric
plate by two powers in the atom-plate separation). It is worth noting
that a similar behaviour, i.e., the same hierarchy of power laws and
the same signs have been found for the vdW force between two atoms
\cite{Sucher68,Boyer69,Farina02} and for the Casimir force between two
semi-infinite half-spaces \cite{Henkel04}. This is illustrated in
Tab.~\ref{tab1}, where the asymptotic power laws found for an atom
interacting with an infinitely thick plate, Eqs.~(\ref{eq24}) and
(\ref{eq25}), and an asymptotically thin plate, Eqs.~(\ref{eq28}) and
(\ref{eq29}), are summarized and compared to those obtainable for the
interactions between two atoms or two half-spaces, respectively.  
%%%%%%%%%%%%%%%  T A B L E %%%%%%%%%%%%%%%%%%%%%%%%%%%%%%%%%%%%%%%%%%%
\begin{table}
\begin{center}
 \begin{tabular}{|c||c|c|c|c|}
\hline
 distance&\multicolumn{2}{c|}{short}&
 \multicolumn{2}{c|}{long}\\ 
\hline
 polarizability&$\mathrm{e}\leftrightarrow\mathrm{e}$
 &$\mathrm{e}\leftrightarrow\mathrm{m}$
 &$\mathrm{e}\leftrightarrow\mathrm{e}$
 &$\mathrm{e}\leftrightarrow\mathrm{m}$\\ 
\hline\hline
 atom $\leftrightarrow$ half space
 &\parbox{5ex}{$$-\frac{1}{z^4}$$}
 &\parbox{5ex}{$$+\frac{1}{z^2}$$}
 &\parbox{5ex}{$$-\frac{1}{z^5}$$}
 &\parbox{5ex}{$$+\frac{1}{z^5}$$}\\
\hline
 atom $\leftrightarrow$ thin plate
 &\parbox{5ex}{$$-\frac{1}{z^5}$$}
 &\parbox{5ex}{$$+\frac{1}{z^3}$$}  
 &\parbox{5ex}{$$-\frac{1}{z^6}$$}
 &\parbox{5ex}{$$+\frac{1}{z^6}$$}\\ 
\hline
 atom $\leftrightarrow$ atom 
 &\parbox{5ex}{$$-\frac{1}{z^7}$$}
 &\parbox{5ex}{$$+\frac{1}{z^5}$$}  
 &\parbox{5ex}{$$-\frac{1}{z^8}$$}
 &\parbox{5ex}{$$+\frac{1}{z^8}$$}\\ 
\hline
 half space $\leftrightarrow$ half space 
 &\parbox{5ex}{$$-\frac{1}{z^3}$$}
 &\parbox{5ex}{$$+\frac{1}{z}$$}  
 &\parbox{5ex}{$$-\frac{1}{z^4}$$}
 &\parbox{5ex}{$$+\frac{1}{z^4}$$}\\ 
\hline
\end{tabular}
\end{center}
\caption{
\label{tab1}
Signs and asymptotic power laws of the forces between
various polarizable objects. In the table heading, $\mathrm{e}$ stands
for a purely electric object, $\mathrm{m}$ denotes a purely magnetic
one. The signs $+$ and $-$ denote repulsive and attractive forces,
respectively.}
\end{table}
%%%%%%%%%%%%%%%%%%%%%%%%%%%%%%%%%%%%%%%%%%%%%%%%%%%%%%%%%%%%%%%%%%%%%%

For weak magnetodielectric properties, i.e., $\chi_\mathrm{e}(iu)$
$\!=$ $\!\varepsilon(iu)$ $\!-1$ $\!\ll$ $\!1$ and
$\chi_\mathrm{m}(iu)$ $\!=$ $\!\mu(iu)$ $\!-$ $\!1$ $\!\ll$ $\!1$,
the similarity of the results shown in Tab.~\ref{tab1} can be regarded
as being a consequence of the additivity of vdW-type interactions. In
fact, in this case (which for a gaseous medium of given atomic species
corresponds to a sufficiently dilute gas) all results of the table can
be derived from the vdW interaction of two single atoms via pairwise
summation. The additivity can explicitly be seen when comparing the
result found for the asymptotically thin plate with that of the
infinitely thick plate. Expanding the vdW potential of an infinitely
thick plate, Eq.~(\ref{eq23}), in powers of $\chi_\mathrm{e}(iu)$ and
$\chi_\mathrm{m}(iu)$, we find that the leading, first-order
contribution is given by
\begin{align}
\label{eq01}
\Delta_1U(z_\mathrm{A})
&= -\frac{\hbar\mu_0}{8\pi^2}
 \int_0^{\infty}\!\!\mathrm{d}u\,u^2 \alpha^{(0)}(iu)
 \int_0^\infty\!\!\mathrm{d}q\,\frac{q}{b}e^{-2bz_\mathrm{A}}
 \nonumber\\
&\hspace{2ex}\times\,\Biggl\{\Biggl[\biggl(\frac{bc}{u}
\biggr)^2-1+\frac{1}{2}
 \biggl(\frac{u}{bc}\biggr)^2\Biggr]\chi_\mathrm{e}(iu)
 \nonumber\\
&\hspace{8ex}-\Biggl[1-\frac{1}{2}\biggl(\frac{u}{bc}\biggr)^2\Biggr]
 \chi_\mathrm{m}(iu)\Biggr\},
\end{align}
while the first-order contribution to the vdW potential of an
asymptotically thin plate, Eq.~(\ref{eq27c}) reads
\begin{align}
\label{eq02}
\Delta_1U^d(z_\mathrm{A})
&=-\frac{\hbar\mu_0d}{4\pi^2}
 \int_0^{\infty}\!\!\mathrm{d}u\,u^2 \alpha^{(0)}(iu)
 \int_0^\infty\!\!\mathrm{d}q\,qe^{-2bz_\mathrm{A}}
 \nonumber\\
&\hspace{2ex}\times\,\Biggl\{\Biggl[\biggl(\frac{bc}{u}\biggr)^2-1
+\frac{1}{2}\biggl(\frac{u}{bc}\biggr)^2\Biggr]\chi_\mathrm{e}(iu)
 \nonumber\\
&\hspace{8ex}-\Biggl[1-\frac{1}{2}\biggl(\frac{u}{bc}\biggr)^2\Biggr]
 \chi_\mathrm{m}(iu)\Biggr\}\,.
\end{align}
Comparison of Eqs.~(\ref{eq01}) and (\ref{eq02}) shows that to
leading order in $\chi_\mathrm{e}(iu)$ and $\chi_\mathrm{m}(iu)$
the vdW potential of an infinitely thick plate is simply the integral
over an infinite number of thin-plate vdW potentials,
\begin{equation}
\label{eq03}
\Delta_1U(z_\mathrm{A})=
\int_{z_\mathrm{A}}^\infty\frac{\mathrm{d}z}{d}\,\Delta_1U^d(z).
\end{equation}

For media with stronger magnetodielectric properties, medium-assisted
correlations prevent vdW-type forces from being additive. This can be
demonstrated by expanding the vdW potentials given by
Eqs.~(\ref{eq23}) and (\ref{eq27c}) to second order in
$\chi_\mathrm{e}(iu)$ and $\chi_\mathrm{m}(iu)$, resulting in 
\begin{align}
\label{eq04}
&\Delta_2U(z_\mathrm{A})
=-\frac{\hbar\mu_0}{8\pi^2}
 \!\int_0^{\infty}\!\!\mathrm{d}u\,u^2 \alpha^{(0)}(iu)
 \!\int_0^\infty\!\!\mathrm{d}q\,\frac{q}{b}\,e^{-2bz_\mathrm{A}}\nonumber\\
&\quad\times\Biggl\{\Biggl[-\frac{1}{2}\biggl(\frac{bc}{u}\biggr)^2
 \!+\frac{1}{4}+\frac{1}{4}\biggl(\frac{u}{bc}\biggr)^2
 \!-\frac{1}{4}\biggl(\frac{u}{bc}\biggr)^4
 \Biggr]\chi_\mathrm{e}^2(iu)
 \nonumber\\
&\qquad+\Biggl[\frac{1}{4}+\frac{1}{4}\biggl(\frac{u}{bc}\biggr)^2
 -\frac{1}{4}\biggl(\frac{u}{bc}\biggr)^4\Biggr]
 \chi_\mathrm{m}^2(iu)
 \nonumber\\
&\qquad+\Biggl[-\frac{1}{2}+\biggl(\frac{u}{bc}\biggr)^2
 -\frac{1}{2}\biggl(\frac{u}{bc}\biggr)^4\Biggr]
 \chi_\mathrm{e}(iu)\chi_\mathrm{m}(iu) 
 \Biggr\}
\end{align}
and
\begin{align}
\label{eq05}
&\Delta_2U^d(z_\mathrm{A})
=-\frac{\hbar\mu_0d}{4\pi^2}
 \!\int_0^{\infty}\!\!\mathrm{d}u\,u^2 \alpha^{(0)}(iu)
 \!\int_0^\infty\!\!\mathrm{d}q\,qe^{-2bz_\mathrm{A}}\nonumber\\
&\quad\times\Biggl\{\Biggl[-\frac{1}{2}\biggl(\frac{bc}{u}\biggr)^2
 +\frac{3}{4}-\frac{1}{4}\biggl(\frac{u}{bc}\biggr)^2
 \Biggr]\chi_\mathrm{e}^2(iu)
 \nonumber\\
&\qquad+\Biggl[\frac{1}{4}-\frac{1}{4}\biggl(\frac{u}{bc}\biggr)^2
 \Biggr]\chi_\mathrm{m}^2(iu)\Biggr\}\,,
\end{align}
respectively. The leading (second-order) correction due to
medium-assisted correlations can be obtained from the vdW potential of
an atom in front of two asymptotically thin plates. Physically, it can
be ascribed to the process of radiation being reflected at the back
(left) plate while aquiring finite phase shifts upon transmission
trough the front (right) plate. The calculation yields (see Appendix
\ref{appB}) 
\begin{align}
\label{eq08}
&\Delta_2U^{dd}(z_\mathrm{A},s)
=-\frac{\hbar\mu_0d^2}{2\pi^2}
 \!\int_0^{\infty}\!\!\mathrm{d}u\,u^2 \alpha^{(0)}(iu)
 \!\int_0^\infty\!\!\mathrm{d}q\,qb\nonumber\\
&\hspace{1ex}\times\,e^{-2b(z_\mathrm{A}+s)}\Biggl\{\Biggl[-\frac{1}{2}
 +\frac{1}{2}\biggl(\frac{u}{bc}\biggr)^2
 \!-\frac{1}{4}\biggl(\frac{u}{bc}\biggr)^4
 \Biggr]\chi_\mathrm{e}^2(iu)
 \nonumber\\
&\quad+\Biggl[\frac{1}{2}\biggl(\frac{u}{bc}\biggr)^2
 \!-\frac{1}{4}\biggl(\frac{u}{bc}\biggr)^4\Biggr]
 \chi_\mathrm{m}^2(iu)
 \nonumber\\
&\quad+\Biggl[-\frac{1}{2}+\biggl(\frac{u}{bc}\biggr)^2
 \!-\frac{1}{2}\biggl(\frac{u}{bc}\biggr)^4\Biggr]
 \chi_\mathrm{e}(iu)\chi_\mathrm{m}(iu) 
 \Biggr\}.
\end{align}
Note that the leading correction due to multiple reflections between
the plates and fractional transparency of the front (right) plate are
of third order in $\chi_\mathrm{e}(iu)$ and $\chi_\mathrm{m}(iu)$, 
and are thus not relevant for the second-order correction considered 
here. Comparing Eqs.~(\ref{eq04}), (\ref{eq05}), and (\ref{eq08}), one
can easily verify that 
\begin{align}
\label{eq09}
\Delta_2U(z_\mathrm{A})&=
 \int_{z_\mathrm{A}}^\infty\frac{\mathrm{d}z}{d} \,\Delta_2U^d(z)
 \nonumber\\
&\quad +\int_{z_\mathrm{A}}^\infty\frac{\mathrm{d}z}{d}
 \int_0^\infty\frac{\mathrm{d}s}{d}\Delta_2U^{dd}(z_\mathrm{A},s).
\end{align}
As a consequence of medium-assisted correlations the coefficients of
the asymptotic power laws in Tab.~\ref{tab1} cannot be related via
simple additivity arguments in general. However, we note from
Tab.~\ref{tab1} that the corrections only change the coefficients of
the asmptotic power laws, not the power laws themselves. 

%%%%%%%%%%%%%%%%%%%%%%%%%%%%%%%%%%%%%%%%%%%%%%%%%%%%%%%%%%%%%%%%%%%%%%

\subsection{Atom between two infinitely thick plates}
\label{sec3.4}

In Secs.~\ref{sec3.1} and \ref{sec3.2} we have seen that for
sufficiently strong magnetic properties a single magnetodielectric
plate can feature a potential wall. This suggests that two such plates
can feature a potential well, where the effect of multiple reflections
between the plates must be taken into account. Let us consider the
simplest case of an atom placed between two identical infinitely thick
magnetodielectric plates which are separated by a distance $d_1$
$\!\equiv$ $\!s$ [$n$ $\!=$ $\!2$, $j$ $\!=$ $\!1$,
$\varepsilon_1(\omega)$ $\!=$ $\!\mu_1(\omega)$ $\!\equiv$ $\!1$,
$\varepsilon_0(\omega)$ $\!=$ $\!\varepsilon_2(\omega)$ $\!\equiv$
$\varepsilon(\omega)$, $\mu_0(\omega)$ $\!=$ $\!\mu_2(\omega)$
$\!\equiv$ $\!\mu(\omega)$]. {F}rom Eq.~(\ref{eq18}) together with
Eqs.~(\ref{eq11})--(\ref{eq15}) it then follows that ($b_0$ $\!=$
$\!b_2$ $\!\equiv$ $b_\mathrm{M}$)
\begin{align}
\label{eq41}
&U(z_{\rm A})
= \frac{\hbar\mu_0}{8\pi^2}
 \int_0^{\infty} \mathrm{d} u \,u^2 \alpha^{(0)}(iu)
 \int_0^\infty\mathrm{d}q\,\frac{q}{b}
\nonumber\\&\hspace{2ex}\times\, 
\Bigl[e^{-2bz_\mathrm{A}}+e^{-2b(s-z_\mathrm{A})}\Bigr] 
 \biggl[\frac{1}{D_1^s}
\frac{\mu(iu)b-b_\mathrm{M}}
 {\mu(iu)b+b_\mathrm{M}}
\nonumber\\&\hspace{6ex} 
- \biggl(1+2\frac{q^2c^2}{u^2}\biggr) \frac{1}{D_1^p}
 \frac{\varepsilon(iu)b-b_\mathrm{M}}
 {\varepsilon(iu)b+b_\mathrm{M}}\biggr] ,
\end{align}
where the coefficients
\begin{align}
\label{eq42}
&D_1^s = 1-\biggl[\frac{\mu(iu)b-b_\mathrm{M}}
 {\mu(iu)b+b_\mathrm{M}}\biggr]^2e^{-2bs}\le 1,
\\
\label{eq43}
&D_1^p = 1-\biggl[\frac{\varepsilon(iu)b-b_\mathrm{M}}
 {\varepsilon(iu)b+b_\mathrm{M}}\biggr]^2e^{-2bs}\le 1 
\end{align}
describe the effect of multiple reflections of radiation between the
two plates, as can be seen from the expansion
\begin{equation}
\label{eq44}
\frac{1}{D_1^\sigma}=\frac{1}{1-r_{1-}^\sigma r_{1+}^\sigma e^{-2bs}}
=\sum_{n=0}^\infty \big(r_{1-}^\sigma e^{-bs}
 r_{1+}^\sigma e^{-bs}\big)^n\, .
\end{equation}
As a consequence of multiple reflections the vdW potential of an atom
between the two plates, Eq.~(\ref{eq41}), can be different from the
sum of two single-plate potentials, Eq.~(\ref{eq23}).

%%%%%%%%%%%%%%%  F I G U R E %%%%%%%%%%%%%%%%%%%%%%%%%%%%%%%%%%%%%%%%%
\begin{figure}[!t!]
\noindent
\begin{center}
\includegraphics[width=\linewidth]{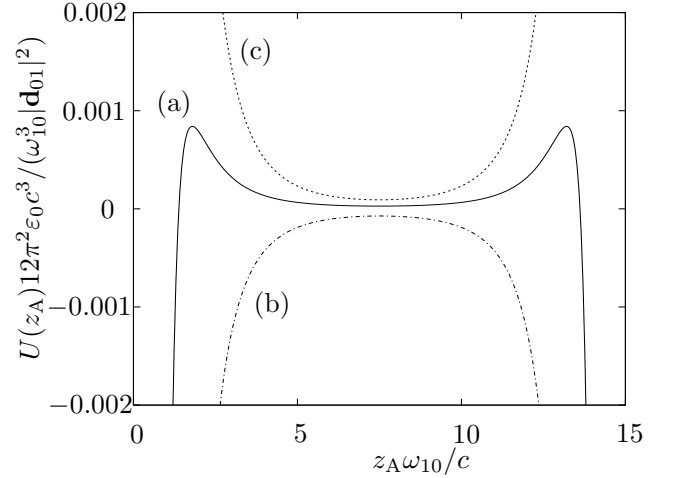}
\end{center}
\caption{
\label{fig7}
The vdW potential of a ground-state two-level atom situated between 
two infinitely thick 
(a) magnetodielectric plates
($\omega_\mathrm{Pe}/\omega_{10}$ $\!=$ $\!0.75$,
$\omega_\mathrm{Te}/\omega_{10}$ $\!=$ $\!1.03$,
\mbox{$\omega_\mathrm{Pm}/\omega_{10}$ $\!=$ $\!2$},
\mbox{$\omega_\mathrm{Tm}/\omega_{10}$ $\!=$ $\!1$},
$\gamma_\mathrm{e}/\omega_{10}$ 
$\!=$ $\!\gamma_\mathrm{m}/\omega_{10}$ $\!=$ $\!0.001$)
(b) dielectric plates [$\mu(\omega)$ $\!\equiv$ $\!1$, other 
parameters as in (a)],
(c) magnetic plates [$\varepsilon(\omega)$ $\!\equiv$ $\!1$, other
parameters as in (a)], which are separated by a distance \mbox{$s$
$\!=$ $\!15c/\omega_{10}$}, is shown as a function of the position of
the atom.
}
\end{figure}
\begin{figure}[!t!]
\noindent
\begin{center}
\includegraphics[width=\linewidth]{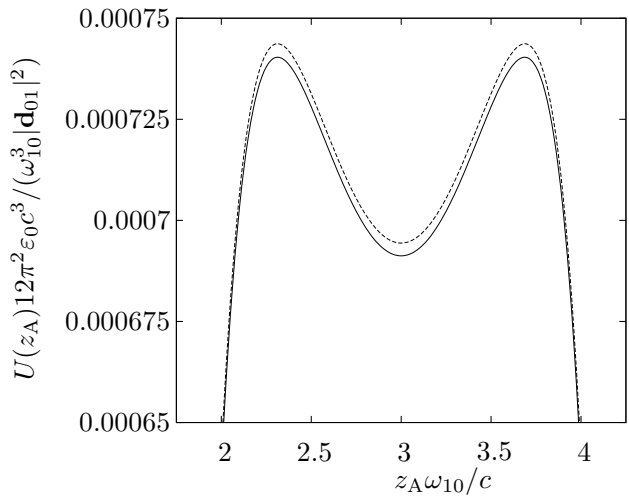}
\end{center}
\caption{
\label{fig8}
The vdW energy of a ground-state two-level atom situated between 
two infinitely thick magnetodielectric plates
($\omega_\mathrm{Pe}/\omega_{10}$ $\!=$ $\!0.75\times 10^5$,
$\omega_\mathrm{Te}/\omega_{10}$ $\!=$ $\!1.03$,
$\omega_\mathrm{Pm}/\omega_{10}$ $\!=$ $\!2\times 10^5$,
$\omega_\mathrm{Tm}/\omega_{10}$ $\!=$ $\!1$,
$\gamma_\mathrm{e}/\omega_{10}$ 
$\!=$ $\!\gamma_\mathrm{m}/\omega_{10}$ $\!=$ $\!0.001$),
which are separated by a 
distance $s$ $\!=$ $\!6c/\omega_{10}$, is shown as a function of the
position of the atom, Eq.~(\ref{eq41}). For comparison, the sum
of two single-plate potentials according to Eq.~(\ref{eq23}) is
also displayed (dashed lines).
}
\end{figure}

%%%%%%%%%%%%%%%%%%%%%%%%%%%%%%%%%%%%%%%%%%%%%%%%%%%%%%%%%%%%%%%%%%%%%%
Examples of the vdW potential  for a two-level atom between two
identical infinitely thick magnetodieletric plates as given by
Eq.~(\ref{eq41}) are plotted in Figs.~\ref{fig7} and \ref{fig8}. In
the case of the parameters chosen in Fig.~\ref{fig7} multiple
reflections are negligible, so that the potentials effectively reduce
to sums of single-plate potentials. This obviously results from the
smallness of the relevant reflection coefficients together with the
relatively large distance between the plates. {F}rom Eqs.~(\ref{eq11})
and (\ref{eq12}) one can easily verify that
\begin{align}
\label{eq45}
&r_{1\pm}^s(iu,q)\le\lim_{q\to\infty}r_{1\pm}^s(0,q)=
\frac{\mu(0)-1}{\mu(0)+1}\, ,\\
\label{eq46}
&r_{1\pm}^p(iu,q)\le\lim_{q\to\infty}r_{1\pm}^p(0,q)=
\frac{\varepsilon(0)-1}{\varepsilon(0)+1}\, .
\end{align}
Hence in the case of the parameters in Fig.~\ref{fig7}(a) we have 
$r_{1-}^sr_{1+}^s$ $\!\le$ $\!0.67$, $r_{1-}^pr_{1+}^p$ $\!\le$
$\!0.044$. In order to demonstrate the effect of multiple reflections,
in Fig.~\ref{fig8} we have (artificially) increased the reflection
coeffiecients, so that almost perfect reflection is realized
($r_{1-}^sr_{1+}^s$ $\!\le$ $\!1$ $\!-$ $\!1\times 10^{-10}$,
$r_{1-}^pr_{1+}^p$ $\!\le$ $\!1$ $\!-$ $\!7.5\times 10^{-10}$),
and reduced the plate separation.
%%%%%%%%%%%%%%%  F I G U R E %%%%%%%%%%%%%%%%%%%%%%%%%%%%%%%%%%%%%%%%%
%%%%%%%%%%%%%%%%%%%%%%%%%%%%%%%%%%%%%%%%%%%%%%%%%%%%%%%%%%%%%%%%%%%%%%
It is seen that multiple reflections lead to a slight lowering of the
vdW potential in the region near the midpoint between the two plates. 

%%%%%%%%%%%%%%%%%%%%%%%%%%%%%%%%%%%%%%%%%%%%%%%%%%%%%%%%%%%%%%%%%%%%%%

\section{Summary and Conclusions}
\label{sec4}

We have studied the problem of the van der Waals force acting on a
ground-state atom in the presence of planar, dispersing, and absorbing
magnetodielectric bodies. Considering an arbitrary planar multilayer
system and restricting our attention to the lowest (nonvanishing)
order of perturbation theory, we have given a general expression for
the vdW potential. The effect of the multilayer system is expressed in
terms of generalized reflection coefficients, which on their part are
determined, inter alia, by the (complex) permittivities and
permeabilities of the layers. Applying the formula to the cases of
an atom being in front of a magnetodielectric plate and between two
such plates, we have placed special emphasis on the competing
attractive and repulsive force components associated with the electric
and magnetic matter properties, respectively. Both numerical and
analytical results are given, the latter referring to some limiting
cases such as the asymptotic behaviour of the potential for thick and
thin plates in the long- and short-distance limits.

In contrast to the well-known attractive vdW force generated by a
purely dielectric plate, a purely magnetic plate leads to a repulsive
force. In the case of genuinely magnetodielectric material, the
influence of the magnetic properties can thus considerably reduce the
strength of the vdW force and---for sufficiently strong magnetic
properties---even create a repulsive potential wall of finite height.
The numerical results show that the height of such a potential wall
sensitively depends not only on the relative strengths of the electric
and magnetic properties, but also on the thickness of the plate. In
particular, they suggest that the maximum height is realized in the
case when the thickness of the plate is comparable to the distance of
the potential maximum from the plate. Comparing the results obtained
for an infinitely thick plate with those found for an asymptotically
thin plate, we have found striking similarities which for weakly
magnetodielectric media can be explained by the additivity of vdW
potentials. Moreover, we have explicitly demonstrated how
medium-assisted correlations lead to a breakdown of additivity for
media with stronger magnetodielectric properties. For an atom being
situated between two magnetodielectric plates each of which features a
potential barrier, a potential well of finite depth can be formed.  
If the plates possess a sufficiently high reflectivity while being
relatively close together multiple reflections can prevent the
resulting potential from being simply the additive superposition of
the two single-plate potentials. 
   
The results show that the advent of artificially made materials with 
controllable magnetodielectric properties will offer novel
possibilities of realizing vdW potentials on demand. The provided
analysis of typical effects relevant for controlling the vdW force in
the case of one and two magnetodielectric plates---namely the
competition between electric and magnetic properties of the material
in the formation of the potential, material absorption, plate
thickness, and multiple reflections---can of course be extended
to more complex multilayer systems by further evaluating the general
formula for the vdW potential of a ground-state atom in planar
multilayer systems. 

%%%%%%%%%%%%%%%%%%%%%%%%%%%%%%%%%%%%%%%%%%%%%%%%%%%%%%%%%%%%%%%%%%%%%%

\begin{acknowledgments}
This work was supported by the Deutsche Forschungsgemeinschaft.
We thank Ho Trung Dung and J.B. Pen\-dry for valuable discussions. 
S.Y.B. is grateful for having been granted a Th\"{u}\-rin\-ger
Landesgraduiertenstipendium and acknowledges support by the E.W.
Kuhlmann-Foundation. T.K. is grateful for being member of
Graduiertenkolleg 567, which is funded by the Deutsche
Forschungsgemeinschaft and the Government of Mecklenburg-Vorpommern. 
\end{acknowledgments}

%%%%%%%%%%%%%%%%%%%%%%%%%%%%%%%%%%%%%%%%%%%%%%%%%%%%%%%%%%%%%%%%%%%%%%

\appendix

%%%%%%%%%%%%%%%%%%%%%%%%%%%%%%%%%%%%%%%%%%%%%%%%%%%%%%%%%%%%%%%%%%%%%%

\section{Long- and short-distance limits}
\label{appA}

The long-distance (short-distance) limit corresponds to separations
$z_\mathrm{A}$ between the atom and the multilayer system which are
much greater (smaller) than the wavelenghts corresponding to typical
frequencies of the atom and the multilayer system. To obtain
approximate results for the two limiting cases, let us analyze the
$u$-integrals in Eqs.~(\ref{eq23}) and (\ref{eq27c}) in a little more
detail and begin with the long-distance limit, i.e.,
\begin{align}
\label{A1}
z_\mathrm{A}\gg 
\frac{c}{\omega_\mathrm{A}^-}\,,\quad 
z_\mathrm{A}\gg \frac{c}{\omega_\mathrm{M}^-}\,,
\end{align}
where $\omega_\mathrm{A}^-$ $\!=$ $\!\mathrm{min}(\{\omega_{k0}|k$
$\!=$ $\!1,2\ldots\})$ is the lowest atomic transition frequency, and 
$\omega_\mathrm{M}^-$ $\!=$
$\!\mathrm{min}(\omega_\mathrm{Te},\omega_\mathrm{Tm})$
is the lowest medium resonance frequency. For convenience, we
introduce the new integration variable $v$ $\!=$ $\!cb/u$
and transform the integral according to
\begin{align}
\label{A3}
\int_0^\infty\mathrm{d}u &
 \int_0^\infty\mathrm{d}q\,\frac{q}{b}\,e^{-2bz_\mathrm{A}}\ldots
 \nonumber\\
&\mapsto\int_1^\infty\mathrm{d}v
 \int_0^\infty\mathrm{d}u\,
 \frac{u}{c}\,e^{-2z_\mathrm{A}vu/c}\ldots\ ,
\end{align}
where $b_\mathrm{M}$ has to be replaced according to 
\begin{equation}
\label{A4}
b_\mathrm{M}\mapsto
\frac{u}{c}\,\sqrt{\varepsilon(iu)\mu(iu)-1+v^2}\,. 
\end{equation}
Inspection of Eqs.~(\ref{eq23}) and (\ref{eq27c}) together with
Eq.~(\ref{A3}) reveals that the frequency interval giving the main
contribution to the respective $u$-integral is determined by a set of
effective cutoff functions, namely
\begin{equation}
\label{C1}
f(u)=e^{-2z_\mathrm{A}u/c},
\end{equation}
\begin{equation}
\label{C2}
g_k(u)=\frac{1}{1+(u/\omega_{k0})^2}\,,
\end{equation}
which enter via the atomic polarizability, cf. Eq.~(\ref{eq3}), and 
\begin{align}
\label{C3}
&h_\mathrm{e}(u)
= \frac{1}{1+(u/\omega_\mathrm{Te})^2}\,,
\\
\label{C4}
&h_\mathrm{m}(u)
= \frac{1}{1+(u/\omega_\mathrm{Tm})^2}\,,
\end{align}
which enter via $\varepsilon(iu)$ and $\mu(iu)$, cf.
Eqs.~(\ref{eq31}) and (\ref{eq32}). The cutoff functions obviously
give their main contributions in regions, where 
\begin{align}
\label{C5}
&\hspace{10ex} u\lesssim \frac{c}{2z_\mathrm{A}} &\hspace{-1ex} \mathrm{for} 
& \quad f(u),\hspace{10ex}\\
\label{C6}
&\hspace{10ex} u\lesssim\omega_{k0} & \hspace{-1ex}\mathrm{for} 
& \quad g_k(u),\hspace{10ex}\\
\label{C7}
&\hspace{10ex} u\lesssim\omega_\mathrm{Te} & \hspace{-1ex}\mathrm{for}
& \quad h_\mathrm{e}(u),\hspace{10ex}\\
\label{C8}
&\hspace{10ex} u\lesssim\omega_\mathrm{Tm} & \hspace{-1ex}\mathrm{for} 
& \quad h_\mathrm{m}(u).\hspace{10ex}
\end{align}
Combining Eq.~(\ref{C5}) with Eq.~(\ref{A1}), we find that the
function $f(u)$ effectively limits the $u$-integration to a
region where
\begin{align}
\label{A5}
&\frac{u}{\omega_{k0}} \le \frac{u}{\omega_\mathrm{A}^-}
\lesssim \frac{c}{2z_\mathrm{A}\omega_\mathrm{A}^-} \ll 1,
\\[1ex]
\label{A5-1}
&\frac{u}{\omega_\mathrm{Te}} \le \frac{u}{\omega_\mathrm{M}^-}
\lesssim \frac{c}{2z_\mathrm{A}\omega_\mathrm{M}^-}\ll 1,
\\[1ex]
\label{A5-2}
&\frac{u}{\omega_\mathrm{Tm}} \le \frac{u}{\omega_\mathrm{M}^-}
\lesssim \frac{c}{2z_\mathrm{A}\omega_\mathrm{M}^-} \ll 1.
\end{align}
Performing a leading-order expansion of the integrands in
Eqs.~(\ref{eq23}) and (\ref{eq27c}) in terms of the small quantities 
$u/\omega_{k0}$, $u/\omega_\mathrm{Te}$, and $u/\omega_\mathrm{Tm}$,
we may set 
\begin{equation}
\label{A6}
\alpha^{(0)}(iu)\simeq\alpha^{(0)}(0),\  
\varepsilon(iu)\simeq\varepsilon(0),\ 
\mu(iu)\simeq\mu(0).
\end{equation}
Combining Eqs.~(\ref{A3}), (\ref{A4}), and (\ref{A6})  
with Eq.~(\ref{eq23}) and Eq.~(\ref{eq27c}), repectively,
and evaluating the remaining $u$-integrals we arrive at
Eq.~(\ref{eq24}) [together with Eq.~(\ref{eq24.1})] and
Eq.~(\ref{eq28}) [together with Eq.~(\ref{eq28.1})]. 

The short-distance limit, on the contrary, is defined by
\begin{align}
\label{B1}
z_\mathrm{A}\ll 
\frac{c}{\omega_\mathrm{A}^+n(0)}\quad\mathrm{and/or}\quad
z_\mathrm{A}\ll \frac{c}{\omega_\mathrm{M}^+n(0)}\,,
\end{align}
where $\omega_\mathrm{A}^+$ $\!=$ $\!\mathrm{max}(\{\omega_{k0}|k$
$\!=$ $\!1,2,\ldots\})$ is the highest inneratomic transition
frequency, $\omega_\mathrm{M}^+$ $\!=\!$ 
$\mathrm{max}(\omega_\mathrm{Te},\omega_\mathrm{Tm})$
is the highest medium resonance frequency, and $n(0)$ $\!=$
$\!\sqrt{\varepsilon(0)\mu(0)}$ is the static refractive index of the
medium. Again, it is convenient to change the integration variables in
Eqs.~(\ref{eq23}) and (\ref{eq27c}), but now we transform according to
\begin{align}
\label{B3}
\int_0^\infty\mathrm{d}u &
 \int_0^\infty\mathrm{d}q\,\frac{q}{b}\,e^{-2bz_\mathrm{A}}\ldots
\nonumber\\
&\mapsto\int_0^\infty\mathrm{d}u\,
 \int_{u/c}^\infty\mathrm{d}b\,
 e^{-2bz_\mathrm{A}}\ldots\ ,
\end{align}
where $b_\mathrm{M}$ has to be replaced according to 
\begin{equation}
\label{B4}
b_\mathrm{M}\mapsto
\sqrt{\frac{u^2}{c^2}\big[\varepsilon(iu)\mu(iu)-1\big] 
 +b^2}\ .
\end{equation} 
Combining Eqs.~(\ref{C6})--(\ref{C8}) with Eq.~(\ref{B1}) reveals that
the functions $g_k(u)$, $h_\mathrm{e}(u)$, and $h_\mathrm{m}(u)$ limit
the $u$-integration to a region where
\begin{equation}
\label{C10}
\frac{z_\mathrm{A}u\sqrt{\varepsilon(iu)\mu(iu)-1}}{c} \lesssim
\frac{z_\mathrm{A}\omega_\mathrm{A}^+n(0)}{c}\ll 1
\end{equation}
and/or
\begin{equation}
\label{C10-1}
\frac{z_\mathrm{A}u\sqrt{\varepsilon(iu)\mu(iu)-1}}{c} \lesssim
\frac{z_\mathrm{A}\omega_\mathrm{M}^+n(0)}{c}\ll 1.
\end{equation}
A valid approximation to the $u$-integrals in Eqs.~(\ref{eq23}) and
(\ref{eq27c}) can hence be obtained by performing a Taylor exansion
in $z_\mathrm{A}u\sqrt{\varepsilon(iu)\mu(iu)-1}/c$. To that end, we
apply the transformation (\ref{B3}) to Eqs.~(\ref{eq23}) and
(\ref{eq27c}), respectively, retain only the leading-order terms in 
$u\sqrt{\varepsilon(iu)\mu(iu)-1}/(cb)$ (which after carrying
out the $b$-integral will yield the leading-order terms in
$z_\mathrm{A}u\sqrt{\varepsilon(iu)\mu(iu)-1}/c$) and obtain
\begin{align}
\label{B6}
&U(z_\mathrm{A}) =
 -\frac{\hbar\mu_0}{8\pi^2}\int_0^\infty\mathrm{d}u\, u^2
 \alpha^{(0)}(iu)
 \int_{u/c}^\infty\mathrm{d}b\,e^{-2bz_\mathrm{A}}
\nonumber\\&\hspace{2ex}\times\,\Biggl(2\frac{c^2b^2}{u^2}\,
 \frac{\varepsilon(iu)-1}{\varepsilon(iu)+1}
 -\biggl\{\frac{\varepsilon(iu)-1}{\varepsilon(iu)+1}
 +\ \frac{\mu(iu)-1}{\mu(iu)+1}
\nonumber\\&\hspace{8ex}+ 
\frac{2\varepsilon(iu)[\varepsilon(iu)\mu(iu)-1]}
 {[\varepsilon(iu)+1]^2}\biggr\}\Biggr) .
\end{align}
and
\begin{align}
\label{B7}
&U(z_\mathrm{A}) =
 -\frac{\hbar\mu_0d}{8\pi^2}\int_0^\infty\mathrm{d}u\, u^2
 \alpha^{(0)}(iu)
 \int_{u/c}^\infty\mathrm{d}b\,be^{-2bz_\mathrm{A}}
\nonumber\\&\hspace{2ex}\times\,\Biggl\{2\frac{c^2b^2}{u^2}\,
 \frac{\varepsilon^2(iu)-1}{2\varepsilon(iu)}
 -\biggl[\frac{\varepsilon^2(iu)-1}{2\varepsilon(iu)}
 \nonumber\\&\hspace{8ex} +\frac{\mu^2(iu)-1}{2\mu(iu)}
 +\frac{\varepsilon(iu)\mu(iu)-1}
 {\varepsilon(iu)}\biggr]\Biggr\}.
\end{align}
After evaluating the  $b$-integrals and keeping only the leading-order
terms in $uz_\mathrm{A}/c$ [note that Eqs.~(\ref{C6})--(\ref{C8})
together with Eq.~(\ref{B1}) imply $uz_\mathrm{A}/c$ $\!\ll$ $\!1$], 
Eq.~(\ref{B6}) and Eq.~(\ref{B7}), respectively, result in
Eq.~(\ref{eq25}) [together with Eqs.~(\ref{eq25.1}) and
(\ref{eq25.2})] and Eq.~(\ref{eq29}) [together
with Eqs.~(\ref{eq29.1}) and (\ref{eq30})].

%%%%%%%%%%%%%%%%%%%%%%%%%%%%%%%%%%%%%%%%%%%%%%%%%%%%%%%%%%%%%%%%%%%%%%

\section{Derivation of Eq.~\ref{eq08}}
\label{appB}

In order to derive Eq.~(\ref{eq08}), we consider the vdW potential of
two plates of thickness $d_1$ $\!\equiv$ $\!d$, $\!d_3$ $\!\equiv$
$\!d'$ which are separated by a distance $\!d_2$ $\!\equiv$ $\!s$ [$n$
$\!=$ $\!j$ $\!=$ $\!4$, $\varepsilon_1(\omega)$ $\!\equiv$
$\!\varepsilon(\omega)$, $\!\varepsilon_3(\omega)$ $\!\equiv$
$\!\varepsilon'(\omega)$,
$\mu_1(\omega)$ $\!\equiv$ $\!\mu(\omega)$,
$\!\mu_3(\omega)$ $\!\equiv$ $\!\mu'(\omega)$,
$\varepsilon_0(\omega)$ $\!=$ $\!\varepsilon_2(\omega)$ 
$\!=$ $\!\varepsilon_4(\omega)$ $\!\equiv$ $\!1$, 
$\mu_0(\omega)$ $\!=$ $\!\mu_2(\omega)$ $\!=$ $\!\mu_4(\omega)$ 
$\!\equiv$ $\!1$]. We assume both plates to be asymptotically thin, 
$\sqrt{\varepsilon(0)\mu(0)}d$ $\!\ll$ $\!z_\mathrm{A}$,
$\sqrt{\varepsilon'(0)\mu'(0)}d'$ $\!\ll$ $\!z_\mathrm{A}$, so that
the inequalities $b_\mathrm{M}d$ $\!\ll$ $\!1$, $b_\mathrm{M}'d'$
$\!\ll$ $\!1$ ($b_1$ $\!\equiv$ $\!b_\mathrm{M}$, $b_3$ $\!\equiv$
$\!b_\mathrm{M}'$) are valid, cf. Eq.~(\ref{eq27b}). Use of
Eqs.~(\ref{eq11}) and (\ref{eq12}) for $l$ $\!=$ $\!n$ $\!=$ $\!4$ and
$l$ $\!=$ $\!3$, followed by a linear expansion in terms of
$b_\mathrm{M}'d'$, yields
\begin{align}
\label{BB1}
&r_{n-}^s \simeq
 \frac{\mu^{\prime 2}(iu)b^2-b_\mathrm{M}^{\prime 2}}{2\mu'(iu)b}\,d'
 +e^{-2bs}r_{2-}^sd'
 \nonumber\\
&\hspace{7ex}\times\,
 \biggl\{1
 -\frac{\mu^{\prime 2}(iu)b^2+b_\mathrm{M}^{\prime 2}}{\mu'(iu)b}
 \nonumber\\
&\hspace{12ex}
+\frac{\mu^{\prime 2}(iu)b^2-b_\mathrm{M}^{\prime 2}}{2\mu'(iu)b}
e^{-2bs}r_{2-}^s\biggr\},\\
\label{BB2}
&r_{n-}^p \simeq
 \frac{\varepsilon^{\prime 2}(iu)b^2-b_\mathrm{M}^{\prime 2}}
 {2\varepsilon'(iu)b}\,d'
 +e^{-2bs}r_{2-}^pd'
 \nonumber\\
&\hspace{7ex}\times\,
 \biggl\{1-
 \frac{\varepsilon^{\prime 2}(iu)b^2+b_\mathrm{M}^{\prime 2}}
 {\varepsilon'(iu)b}
 \nonumber\\
&\hspace{12ex} 
 +\frac{\varepsilon^{\prime 2}(iu)b^2-b_\mathrm{M}^{\prime 2}}
 {2\varepsilon'(iu)b}
 e^{-2bs}r_{2-}^p\biggr\},
\end{align}
while use of the same equations for $l$ $\!=$ $\!2$ and $l$ $\!=$
$\!1$ together with $\!r^s_{0-}$ $\!=$ $\!r^p_{0-}$ $\!=$ $\!0$
leads to, upon linearly expanding in terms of $b_\mathrm{M}d$,
\begin{align}
\label{BB3}
&r_{2-}^s \simeq
 \frac{\mu^{2}(iu)b^2-b_\mathrm{M}^2}{2\mu(iu)b}\,d,
 \\
\label{BB4} 
&r_{2-}^p \simeq
 \frac{\varepsilon^{2}(iu)b^2-b_\mathrm{M}^2}
 {2\varepsilon(iu)b}\,d.
\end{align}
Substituting Eqs.~(\ref{BB3}) and (\ref{BB4}) into Eqs.~(\ref{BB1})
and (\ref{BB2}), respectively, and neglecting terms which are
quadratic in $b_\mathrm{M}d$, we arrive at
\begin{align}
\label{BB5}
&r_{n-}^s \simeq
 \frac{\mu'^2(iu)b^2-b_\mathrm{M}'^2}{2\mu'(iu)b}\,d'
 +\frac{\mu^{2}(iu)b^2-b_\mathrm{M}^{2}}
 {2\mu(iu)b}\,e^{-2bs}d
 \nonumber\\
&\hspace{10ex}\times\,
 \biggl[1-
 \frac{\mu'^2(iu)b^2+b_\mathrm{M}'^2}{\mu'(iu)b}\,d'\biggr]\,,\\
\label{BB6}
&r_{n-}^p \simeq
 \frac{\varepsilon'^2(iu)b^2-b_\mathrm{M}'^2}{2\varepsilon'(iu)b}\,d'
 +\frac{\varepsilon^{2}(iu)b^2-b_\mathrm{M}^{2}}
 {2\varepsilon(iu)b}\,e^{-2bs}d
 \nonumber\\
&\hspace{10ex}\times\,
 \biggl[1-
 \frac{\varepsilon'^2(iu)b^2+b_\mathrm{M}'^2}
 {\varepsilon'(iu)b}\,d'\biggr].
\end{align}

The leading correction due to medium correlations can be extracted
from Eqs.~(\ref{BB5}) and (\ref{BB6}) by retaining only the two-plate
contribution, i.e., the term which is linear in both $b_\mathrm{M}d$
and $b_\mathrm{M}'d'$. We expand the result up to linear order in 
$\chi_\mathrm{e}(iu)$, $\chi_\mathrm{m}(iu)$, $\chi_\mathrm{e}'(iu)$
$\!\equiv$ $\!\varepsilon'(iu)-1$, and $\chi_\mathrm{m}'(iu)$
$\!\equiv$ $\!\mu'(iu)-1$, thereby discarding terms which are
independent of $\chi_\mathrm{e}'(iu)$ and $\chi_\mathrm{m}'(iu)$,
leading to
\begin{align}
\label{BB7}
&r_{n-}^s\simeq
 b^2d^2e^{-2bs}\Biggl\{
 \frac{1}{2}\biggl(\frac{u}{bc}\biggr)^4\chi_\mathrm{e}^2(iu)\!
 -\!\!\Biggl[\biggl(\frac{u}{bc}\biggr)^2
 \!\!\!-\frac{1}{2}\biggl(\frac{u}{bc}\biggr)^4\Biggr]
 \nonumber\\
&\hspace{2ex}\times\,
 \chi_\mathrm{m}^2(iu)
 -\!\Biggl[\biggl(\frac{u}{bc}\biggr)^2
 \!-\biggl(\frac{u}{bc}\biggr)^4\Biggr]\chi_\mathrm{e}(iu)
 \chi_\mathrm{m}(iu)\Biggr\}\,,\\
\label{BB8} 
&r_{n-}^p\simeq
 b^2d^2e^{-2bs}\Biggl\{
 -\!\Biggl[\biggl(\frac{u}{bc}\biggr)^2
 \!-\frac{1}{2}\biggl(\frac{u}{bc}\biggr)^4\Biggr]
 \chi_\mathrm{e}^2(iu)
 \nonumber\\
&+\frac{1}{2}\biggl(\frac{u}{bc}\biggr)^4\!\chi_\mathrm{m}^2(iu)\!
 -\!\!\Biggl[\biggl(\frac{u}{bc}\biggr)^2
 \!\!\!-\biggl(\frac{u}{bc}\biggr)^4\Biggr]\chi_\mathrm{e}(iu)
 \chi_\mathrm{m}(iu)\!\Biggr\}\,,
\end{align}
where we have set $d'$ $\!=$ $\!d$, $\chi_\mathrm{e}'(iu)$ $\!=$
$\!\chi_\mathrm{e}(iu)$, and $\chi_\mathrm{m}'(iu)$ $\!=$
$\!\chi_\mathrm{m}(iu)$. Substitution of Eqs.~(\ref{BB7}) and
(\ref{BB8}) into Eq.~(\ref{eq19}) leads to Eq.~(\ref{eq08}). In
order to see that this leading correction corresponds to the process
of radiation being reflected at the back (left) plate while aquiring
finite phase shifts upon transmission trough the front (right) plate,
we note that up to linear order in $\chi_\mathrm{e}'(iu)$,
$\chi_\mathrm{m}'(iu)$, and $d'$, the terms in square brackets in
Eqs.~(\ref{BB5}) and (\ref{BB6}) are equal to the phase factor
$e^{-2b_\mathrm{M}'d'}$, as can be easily verified by recalling
Eq.~(\ref{eq8}):
\begin{align}
\label{BB9} 
&1-\frac{\mu'^2(iu)b^2+b_\mathrm{M}'^2}{\mu'(iu)b}\,d'\nonumber\\
&\hspace{2ex}\simeq
1-2b\Biggl\{1+\frac{1}{2}\biggl(\frac{u}{bc}\biggr)^2
 \big[\chi_\mathrm{e}(iu)+\chi_\mathrm{m}(iu)\big]\Biggr\}d'
 \nonumber\\
&\hspace{2ex}\simeq 1-2b'_\mathrm{M}d'\simeq e^{-2b_\mathrm{M}'d'}, 
\end{align}
similar for Eq.~(\ref{BB6}).

%%%%%%%%%%%%%%%%%%%%%%%%%%%%%%%%%%%%%%%%%%%%%%%%%%%%%%%%%%%%%%%%%%%%%%

\end{document}